\documentstyle[multicol,epsfig,aps,pre,array]{revtex}
\begin{document}
\draft

\title{Morphological Instability and Dynamics of Fronts in Bacterial Growth
Models with Nonlinear Diffusion}

\author{Judith M\"uller\thanks{Present address: DigitalDNA Laboratories, Motorola, Austin, TX 78721, U.S.A. } and Wim van Saarloos}
\address{Instituut--Lorentz, Universiteit Leiden, Postbus 9506,
  2300 RA Leiden, The Netherlands}
\date{\today} \maketitle

\begin{abstract}    
Depending on the growth condition bacterial colonies can exhibit different
morphologies. As argued by Ben-Jacob {\em et al.} there is biological 
and modeling evidence that a non-linear diffusion coefficient of the type 
$D(b) = D_0 b^{k}$ is a basic mechanism which underlies
almost all of the patterns and which generates a long wavelength instability. We study a 
reaction-diffusion system with a non-linear diffusion coefficient and find 
that a unique planar traveling front solution exists
whose velocity is uniquely determined by $k$ and $D = D_0/D_n$ where $D_n$
is the diffusion coefficient of the nutrient. Due to the fact that the 
bacterial diffusion coefficient vanishes when $b \rightarrow 0$, in the front 
solution $b$ vanishes in a singular way. As a result the standard linear 
stability analysis for fronts cannot be used. We introduce an extension of the
stability analysis which can be applied to singular fronts, and use the method
to perform a linear stability analysis of the planar bacteriological growth 
front. We show that a non-linear diffusion coefficient generates a long 
wavelength instability for $k>0$
and $D<D_c(k)$. We map out the region of stability in the $D$-$k$-plane
and determine the onset of stability which is given by $D_c(k)$. 
Both, for $D \rightarrow 0$ and $k \rightarrow \infty$  the dynamics
of the growth zone 
essentially reduces to that of 
 a sharp interface problem which is reminiscent of a so-called one-sided
 growth problem where the growth velocity is proportional to
 the gradient of a diffusion field ahead of the interface. The moving boundary approximation
 that we  derive in these limits is quite accurate but surprisingly does not
 become a proper asymptotic theory in the mathematical strict sense
in the limit $D\to 0$, due to lack of
full separation of scales on all dynamically relevant
length scales.   Our linear stability 
 analysis and sharp interface formulation will also be applicable  to other examples of interface
 formation due to nonlinear diffusion, like in  porous media
 or in the problem of vortex motion in superconductors.

\end{abstract} 
\pacs{PACS numbers: 5.40+j, 5.70.Ln, 61.50.Cj}

\begin{multicols}{2}

% introduction
\section{Introduction}
\subsection{Background of the Problem}
Recently the growth of bacterial colonies under different growth conditions
has been the focus of attention of several groups in the physics community 
since it exhibits different elaborate branching patterns. For an extensive
review and entrance to the literature, see
\cite{shapiro,benjacob,andras}. Already in  
1989,  Fujikawa and  
Matsushita \cite{fujikawa} stressed that bacterial colonies could grow
patterns similar to the type known from the study of physical systems such as 
diffusion-limited aggregation.
A complete morphology diagram  has been obtained for the colonies
of Bacillus subtilis \cite{shapiro,rafols,wakita}, 
where the important control parameter are the agar 
concentration which influences the diffusion of the bacteria as well as of the
nutrient, and the initial nutrient concentration. It includes some 
interesting regimes such as diffusion limited aggregation, dense branching
morphologies, Eden-like and ring patterns. Although the visual appearance
of the patterns is very similar to those of physical systems, at the
microscopic level their  
growth mechanism has to be different --- the question then becomes 
whether or not these microscopic differences affect 
the overall large-scale pattern dynamics.  For instance,  the building units are bacteria
which are themselves micro-organisms and thus living systems. To survive
they have to cope with hostile environmental conditions
which made them develop quite sophisticated cooperation mechanisms and
communication skills, such as direct cell-cell interaction via extra-membrane
polymers, collective production of extra-cellular ``wetting'' fluid for
movement on hard surfaces, long-range chemical signaling, such as quorum
sensing and chemotactic signaling, just to name a few. Different models have 
been proposed which include one or several of these mechanisms, and are able 
to reproduce the rich morphology diagram quite well. Instead of exploring
the richness and diversity of the behavior of bacterial colonies, we want to
concentrate on the basic mechanism which underlies all these patterns.
Since they appear as an interface separating  a region occupied by
the bacteria from a bacteria-free region which 
propagates as the colony is expanding, we look for an interface model 
which includes a long wavelength instability. Although these models have
been developed and studied for pattern-forming, non-living systems
such as crystal growth \cite{langer,benjacob1,benjacob2},  where a sharp interface
formulation is well justified, even at quite small length scales, here 
the existence of interface-type fronts is not obvious   from the start,
but is something that should emerge from the continuum
equations describing the dynamics..
Reaction-diffusion type  models with a non-linear diffusion coefficient for
the bacteria density have been argued to be a good candidate for being the proper
starting point to analyze
the instability mechanism since they were able to reproduce many aspects of 
the above mentioned morphology diagram
\cite{benjacob,benjacob3,kozlovsky,lac,mimura}.

The biological motivation that has been proposed for  non-linear
diffusion coefficient is the  
way bacteria move. Although there are different ways of moving we are 
interested in bacteria which swim by 
propelling themselves with their flagella in straight lines and change their 
direction in a random fashion by tumbling which can be described by a random 
walk. However, for the propelling mechanism to work a liquid with low 
viscosity is required. Since bacteria by themselves are able to secret 
this liquid, their presence is required to generate the lubricant layer 
necessary for diffusion. This behavior can be captured qualitatively by a 
bacteria density dependent diffusion coefficient as has been proposed
in particular by Ben-Jacob and co-workers in 
\cite{kozlovsky}. A consequence of it is that the branches of bacterial
colonies are confined by a sharp envelope which is supported by the 
observation with optical microscopes \cite{benjacob,kozlovsky}.

However, we would like to not that the arguments supporting
a non-linear diffusion model are still not conclusive and
more of a qualitative nature. In addition it is clear that
it does not appear to be relevant for the growth patterns at
large agar
concentrations where  the bacteria are non-motile \cite{wakita}, and
the relevance for the regions where bacteria are motile is still under
discussion. In this
paper we will not address the question of the biological relevance of
the model; instead we aim to contribute to the debate by working out
 the stability diagram and uncovering its essential dynamics. An
additional non-biological contribution of our paper is that we introduce
new methods to deal mathematically with singular fronts. As we discuss
below, this is likely to have implications in other subfields of physics.

In passing, we also note that  it has been shown recently \cite{kessler}
that if one 
extends the model by introducing an effective cutoff in the reaction
term modelling the bacterial growth, while keeping the bacterial
diffusion term linear, one also recovers the type of front instability
necessary to understand bacterial patterns. The motivation for such a
cutoff would be simply the fact that bacteria are
discrete entities, so that at some small density a continuum
formulation breaks down.
 Both mechanism (nonlinear diffusion and discrete entity
cutoff effects to continuum formulations)  are not mutually exclusive
and can
be operative simultaneously, but the detailed studies of various models by a number of authors
 \cite{benjacob,benjacob3,kozlovsky,lac,mimura} suggests that the
nonlinear diffusion mechanism is  the most important  one of the two
\cite{note1}. 

We concentrate on the effect of a nonlinear diffusion coefficient
here since  in spite of the  suggestion that a nonlinear
diffusion coefficient 
is a possible mechanism to generate the complex morphology
diagram,  a clear understanding of this instability mechanism is still
missing.
This is surprising since also from a mathematical point of view it is an interesting 
problem as it defines a new class of fronts which do show up in other 
systems with density dependent diffusivity, such as porous media
\cite{bert,bear,barenblatt}. Furthermore, magnetic flux vortices in
superconductors \cite{gilchrist,surdeanu}. Clearly, understanding the
similarities and differences between instabilities in magnetic flux
patterns and the  well-studied Mullins-Sekerka  
instability mechanism is clearly of importance. Considering the amount
of work and attention there has  
been in the recent years to understand the mechanisms behind bacterial colony 
growth it might at first sight seem surprising that not even a
stability analysis of planar fronts solutions has been  
performed. An important  reason for this  is that as the problem
involves singularities: these make
the standard stability calculations break down, so new techniques
have to be introduced to even perform the linear stability analysis. 
We have been able to resolve the problem and thus  perform an explicit linear
stability analysis of planar fronts which allows us to determine the
regions of stability 
in parameter space. Our extension of the standard stability
calculation is not limited  
to the particular bacterial growth problem we focus on here. Instead it should
be applicable to a large class of growth problems with singular fields, e.g.,
other problems which involve  nonlinear diffusion,
like the vortex
patterns in superconductors \cite{gilchrist,surdeanu} just mentioned, should be
amenable to the same type of  
analysis.

In some limits, in particular in the limit that the bacterial
diffusion coefficient becomes much smaller than the one of the
nutrient, the fronts in the models that have been studied become
rather sharp. A second important question therefore is to what extent
a moving boundary approximation, in which the front is viewed as a
mathematically sharp
interface on the scale of the patterns, becomes appropriate --- such
approximations are often very helpful for analyzing pattern forming
problems (see e.g. \cite{karma} for an application to dendritic growth
and an entry into the vast ``phase field model'' literature). Some
steps in this direction for the bacterial growth problem were taken by
Kitsunezaki \cite{kits}. We
address this question in more detail in this paper and, quite
remarkably, find that 
while in the limit of small bacterial diffusion a moving boundary
approximation is  quite accurate it does not emerge as the lowest
order description in a mathematically well-defined limit. The reason
for this is that even for small diffusion, the dynamically relevant
length scales (i.e., those corresponding to unstable modes in the
linear stability calculation of planar fronts) are not all large in
comparison with the front width. This result shows that bacterial
growth problems  with nonlinear diffusion of the type encountered in
the porous media equation \cite{bert,barenblatt} are mathematically in some crucial ways
different from the standard type of growth problems. Physically, their 
dynamics is closest to those of the so-called one-sided growth
problems \cite{langer}.

\subsection{The Model}
Since we would like to concentrate on the basic mechanism which generates 
a long-wavelength instability we confine our analysis to the most
basic model of Ben-Jacob {\em et al.} \cite{benjacob3,kozlovsky},
namely a two-dimensional reaction-diffusion 
system for the bacteria 
density $b({\bf r},t)$ with a nonlinear diffusion term, and the
nutrient density  
$n({\bf r},t)$ with a linear diffusion term:
\begin{eqnarray} 
{{\partial b }\over{\partial t}} &  = & \nabla D(b) \nabla b + f(n,b), \label{modelbg}  \\
{{\partial n }\over{\partial t}} &  = & D_n \nabla^2 n - g(n,b), \label{modelng} 
\end{eqnarray}
with $D_n$ describing the diffusion constant of the nutrient, and
\begin{equation} 
D(b) = D_0 b^k
\end{equation}
implying a bacteria density dependent diffusion coefficient 
as was motivated before. For simplicity we assume the following reaction term
\begin{equation} 
f(n,b) = g(n,b) = n b, \label{reaction}
\end{equation} 
which  in chemical terms is like a bilinear auto-catalytic reaction:
\begin{equation}
N + B \rightarrow 2 B.
\end{equation}
Biologically it models that the bacteria $B$ eat  a nutrient $N$ to
duplicate themselves.
 This involves a conservation law and is clearly an oversimplification,
since part of the 
energy is also used for movement and other metabolic activities. For
the growth process we want 
to study here, this should not matter. For the same reason,  we also leave out 
in this paper another biologically important feature, sporulation, a
transition of motile  
bacteria into a stationary state; this occurs if there is a deficiency of 
nutrient, which 
seem to play an important role in the later stage of the branching process.
During sporulation bacteria stop normal activity such as movement and use all 
their internal reserves to metamorphose from an active motile cell to a spore,
a sedentary durable ``seed'' which is immotile and hence cannot participate
to the diffusion process. The sporulation process can be included in the model
by adding a term $-\mu b$ on the right side of (\ref{modelbg}). Although the 
simulation by Kitsunezaki \cite{kits} indicate that this death term does affect
the stability of planar fronts, we will not take it into account here since 
the most crucial ingredient is the nonlinear diffusion coefficient of $b$ as 
it assumes that without bacteria there is no diffusion. As we will see
this implies a front profile which goes abruptly to zero, 
 with a divergent slope for $k > 1$. This characteristic is supported by 
experimental
observations of some kinds of bacteria, where one observes a clearly defined 
envelope  (such a comparison suggests a value of $k$ of about one). 
The question we want to study now, is whether this kind of
diffusion is enough to generate a long wavelength instability. It should be
noted here, that for $k=0$ the system has been studied by 
\cite{scott,horvath,toth}.
They showed that  bilinear autocatalysis alone is not sufficient to 
destabilize a planar front. Only in the presence of an  
 autocatalysis term proportional to $b^{\gamma}$ with $\gamma>1$ and $D_n > \beta_c D_0$ where $\beta_c$ depends on
the amount and order of autocatalysis a planar front is unstable toward long 
wavelength 
perturbations \cite{note1,note2,note2a}. Thus, any instability we observe for $k>0$ is due to the
nonlinearity in the diffusion term.
By rescaling the diffusion constant $D=D_0/D_n$ and replacing $f(n,b)$ and
$g(n,b)$ by (\ref{reaction}), we obtain the following nonlinear 
reaction-diffusion system:
\begin{eqnarray} 
{{\partial b }\over{\partial t}} &  = & \frac{D}{k+1} \nabla^2 b^{k+1} + n b, 
\label{modelb} \\
{{\partial n }\over{\partial t}} &  = & \nabla^2 n - n b, \label{modeln} 
\end{eqnarray} 
which contains two parameters, $D$ the rescaled diffusion constant, and $k$
describing the nonlinearity and the stiffness of the front --- in
writing the above equations, we have used the freedom to 
choose appropriate time and length scales, and to rescale the fields
$n$ and $b$ appropriately to set all other prefactors equal to one. We
will be interested in front solutions of this equation where far ahead
of the front the nutrient field $n \to 1$; as we will discuss in more
detail below,  this asymptotic value is
also immaterial, as the problem with another asymptotic value can be
rescaled to our problem with a renormalized value of $D$.

A nonlinear diffusion behavior like in (\ref{modelb}) also arises in the
so-called porous media equation \cite{bert,bear,barenblatt}. There is a vast literature on this
equation \cite{bert,barenblatt};
for us, the essential feature is that it gives rise to moving front solutions
with compact support, i.e., for which the field $b$ is  zero in some
regions of space. At the point where $b$ vanishes, it does so in a singular
way, and this invalidates  the usual linear stability analysis.

\subsection{Overview  of methods and results} 
For the reader not interested in the mathematical details of the derivation,
we now summarize the main results of the analysis.
The model (\ref{modelb})-(\ref{modeln}) has two homogeneous states:
a stable solution $(c_b,0)$ in which only bacteria are present, and an 
unstable solution $(0,c_n)$ with only nutrient. Thus, we can study the 
propagation of the stable state $(c_b,0)$ into the unstable one $(0,c_n)$, 
implying for our system
the propagation of the bacteria field into the nutrient field. To study such a
propagation we look for one-dimensional traveling front solutions which appear
for a system with initial conditions in which the system is in the unstable 
state and a small perturbation at $x \rightarrow -\infty$ starts to invade it.
Assuming that the front propagates with a steady velocity $v$, we can 
reformulate the model in a co-moving frame which reduces 
(\ref{modelb})-(\ref{modeln}) to a one dimensional system of ODE's which is 
much easier to analyze. Its solution will be found numerically by a shooting
method as will be explained in section~II. 

\begin{figure}[tb]
\begin{center}
\leavevmode
\psfig{figure=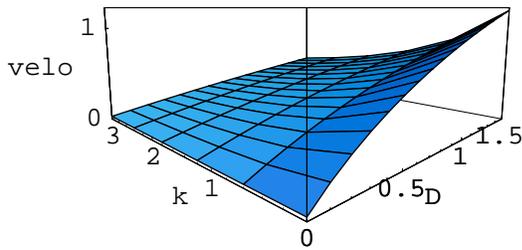,width=7cm}
\end{center}
\narrowtext
\caption{The  front velocity as a function of $D$ and $k$, as
  determined from the analysis in section II. }
\label{figvelo}
\end{figure}

We find that there generally  is a clearly defined unique reaction front,
of which $b$ vanishes with a diverging slope for $k>1$ 
(see Figs.~\ref{figDprofiles} and \ref{figkprofiles}
below). The qualitative features of these fronts are 
consistent with the earlier simulation results of Ben-Jacob {\em et al.} \cite{benjacob3,kozlovsky}
 and can be traced back to the nonlinear diffusion. The characteristic singular 
behavior of the front makes the study of the problem mathematically and 
numerically challenging and intriguing.
The solution provides us with a unique velocity, which depends on
$D$ and $k$ and is shown in Fig.~\ref{figvelo}. More detailed plots of
the behavior  of the velocity as a function of $D$ and $k$ are
presented later  in section II of  this paper.
% \begin{figure}[tb]
% \begin{center}
% \leavevmode
% \psfig{figure=profiles.k2.d.1.eps,width=7cm}
% \end{center}
% \narrowtext
% \caption{Uniformly translating front profiles of bacteria and nutrient densities for $k=2$ and 
% $D=0.3$.}  
% \label{figprof}
% \end{figure}
\begin{figure}[tb]
\begin{center}
%\leavevmode
\vspace*{-1.cm}

\psfig{figure=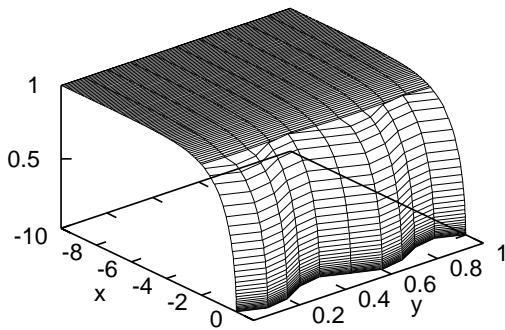,width=8cm}
\end{center}
\vspace*{-0.5cm}

\narrowtext
\caption{Perturbed front profiles of bacteria densities. The front
propagates into the $x$-direction, and has a sinusoidal modulation in
the $y$-direction.}
\label{figprof3d}
\end{figure}
In order to study the stability of the front which is the content of 
section~III
we have to perturb the planar front. Due to the singular behavior of the 
planar front a perturbation of the front is not only a simple perturbation 
in the fields $b$ and $n$ but also in the geometry of the front as sketched 
in Fig.~\ref{figprof3d}.

\begin{figure}[tb]
\begin{center}
\leavevmode
\psfig{figure=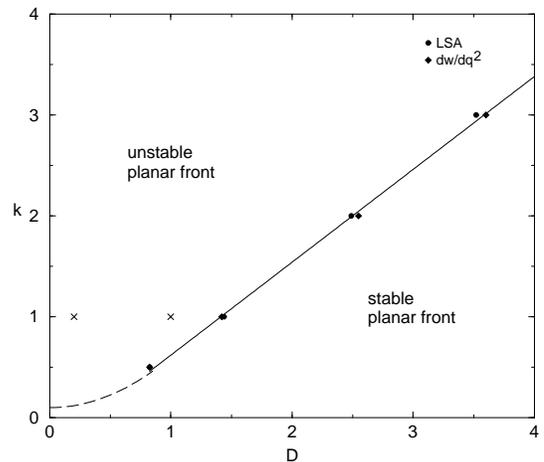,width=7cm}
\end{center}
\narrowtext
\caption[]{Stability diagram for parameters $D$ and $k$. Filled circles show
where the region of stability of planar fronts starts as determined by a
numerical linear stability analysis, filled diamonds show the same boundary
as obtained from the solvablity formula for $d^2\omega /dq^2|_{q=0}$
derived in section~III.C. For $k=0.5$ 
both methods give the same value up to the size of the symbol. The solid line
is there to guide the eye, the dashed line hints at the fact that while we
expect the line of $D_c(k)$ to approach the origin we do not know the precise
analytic behavior of $D_c(k)$ for $k \rightarrow 0$, since for $k=0$ the 
planar front is stable for all $D$ \cite{horvath,scott}. The two crosses
represent the simulation performed by Kitsunezaki \cite{kits}. 
For $D=0.2$ the front in these simulations was unstable which is
consistent with our analysis.  
For $D=1.0$ the planar front was stable, which does not agree with our 
analysis. The probable cause of this apparent discrepancy  is discussed in
the main text.}
\label{figonset}
\end{figure}

Our stability analysis  implements the idea that a proper
Ansatz consists of two contributions, a perturbation in the line of the 
singular front, and the perturbation in the fields away from the singular 
line. Both these contributions have to be determined self-consistently. 
For $k>0$ we observe that for $D<D_c(k)$ the planar front is unstable and
has a long wavelength instability. Thus, a nonlinear diffusion coefficient
together with a bilinear autocatalysis-type reaction term are
sufficient to generate a  
long wavelength instability. For $D>D_c(k)$ the planar front is linearly 
stable. Hence, in the $D$-$k$-parameter space there exist regions of stability
and instability of a planar front. 
We determine these regions in two different ways, one by 
performing numerically a linear stability analysis (LSA) as is done in 
section~III.A, the other by an expansion for small growthrate $\omega$
and wavenumber $q$ around the planar 
profile as is done in section~III.C. 
Fig.~\ref{figonset} shows the stability diagram as a function of $D$ and
$k$. 

Filled circles show the onset of the region of stability of planar 
fronts as determined by a numerical linear stability analysis, 
filled diamonds show the same boundary as obtained from the exact
expression for $d^2\omega /dq^2|_{q=0}$ derived in section~III.C.
Both methods give results which are in very good agreement with each
other, as they should.
The solid line is there to guide the eye, the dashed line hints to the fact 
that while we expect the line of $D_c(k)$ to approach the origin we do not 
know the precise analytic behavior of $D_c(k)$ for $k \rightarrow 0$, since 
for $k=0$ the planar front is stable for all $D$ \cite{horvath,scott}.
We will not analyse the precise behavior in the limit $k \rightarrow 0$
in detail, both because it does not appear to be of practical relevance, and
because the model is very sensitive to slight changes in this limit:
an effective cut-off which arises for discrete particle effects turns the 
model weakly unstable\cite{kessler}, but   a continuum model with 
a different reaction term has the same effect. In particular, if we change the reaction term
$nb$ in (\ref{modelb}), (\ref{modeln}) to $n b^\gamma$, then for any 
$\gamma > 1$ we expect for the limit $k \rightarrow 0$ $D_c$ to be
finite; in other words, for $ \gamma>1$ the stability boundary crosses
the $D$-axis at a nonzero value of $D$.
For $\gamma = 2$, it is in fact known that $D_c(k=0) \approx 0.34$ 
\cite{horvath}.

 The two crosses in Fig.~\ref{figonset} represent the simulation performed by 
Kitsunezaki \cite{kits}. Whereas for $D=0.2$ his planar front was unstable 
which is consistent with our analysis, his planar front for $D=1.0$
appeared to be 
stable, in apparent contradiction  with our results. However, the
simulations were done for a system of width $L_y=40$ and up to time 
$t=200$. From our results for 
the dispersion relation for $k=1$ and $D=1$ which is very  similar to
the one  shown  in Fig.~\ref{figdispA} in section III, we find that
the characteristic length scale of the  fastest 
growing mode  is
$L_m \approx 31$, while the   
associated characteristic time for this fastest growing mode is approximately
$t_m = 520$. Hence, it is likely that the system width is too small and
the simulation time too short to observe the instability. It would
therefore be useful to
redo the simulation for a bigger system and longer times. In fact,
this illustrates the difficulty  of using simulations alone to study
the systems, especially if only a few parameter values can be studied
over a limited time range and system size. On the other hand, our
explicit stability 
analysis allows us to map out the phase diagram in a relatively straightforward way.

In section~IV we map the system with a moving boundary approximation
to a sharp interface problem guided by the success this approach had in
analyzing and understanding the Mullins-Sekerka instability mechanism \cite{langer},
the long wavelength instability associated very generally with
diffusion-limited or Laplacian growth processes. We obtain by a
multiscale expansion equations for 
$b$ and $n$ which are valid in the outer bulk fields, and which are connected 
by boundary conditions. The boundary conditions are obtained by using
solvability type  arguments to integrate out the internal degrees of
freedom of the inner reaction region.  As was already mentioned before,
the moving boundary approximation is closest to the so-called
one-sided growth models and is quite accurate for small $D$, but
it never becomes mathematically correct in the limit $D\to 0$ for all
dynamically relevant length scales.

\section{Planar Front}
% planar front
There exist two trivial homogeneous solutions: 
$n(x,t) = c_n, b(x,t) = 0$, which implies some constant food level and no 
bacteria. This state is unstable since any amount of bacteria will be enough
to let the bacteria density grow. The other trivial homogeneous state is 
$n(x,t) = 0, b(x,t) = c_b$, which assumes a constant bacteria 
density and no food. This state is stable in the present model without sporulation.
% As has been discussed already the bacteria 
% can survive under austere conditions such as no food for a very long time by 
% evolving into a spore state in which their metabolism is severely reduced. 
In addition there exist a steady-state solution in which the stable state
$(c_b,0)$ propagates with a constant velocity $v$ into the unstable state 
$(0,c_n)$, implying the propagation of the bacteria field into the nutrient 
field. Starting from an initial condition
in which the unstable nutrient state is perturbed by a small amount of bacteria 
at the left, the bacteria field invades the nutrient state in the
form of a well defined reaction front propagating to the right. 
Since we are first interested in a planar front, we can restrict ourselves to 
one dimension. To obtain the uniformly translating front solution it is 
convenient to express the reaction-diffusion system in a co-moving frame in
which the new coordinate $\xi$ travels with the  velocity $v_0$ of
the front, 
$\xi = x - v_0 t$. The temporal derivative then transforms as 
$\partial_t|_x = \partial_t|_\xi - v_0 \partial_{\xi}|_t$. For a front 
translating with uniform velocity $v_0$, the explicit time derivative vanishes 
and (\ref{modelb})-(\ref{modeln}) reduces to:
\begin{eqnarray} 
\frac{D}{k+1} \frac{d^2 b^{k+1} }{d \xi^2} 
+ v_0 \frac{d b}{d \xi} + n b &  = & 0,\label{model1db}  \\
\frac{d^2 n}{d \xi^2} + v_0 \frac{d n}{d \xi} - n b 
&  = & 0.\label{model1dn}
\end{eqnarray} 
This is a system of two ODE's of second order. The boundary conditions at
$\xi \rightarrow \pm \infty$ are given by the two homogeneous states. 
By choosing a right-moving front we obtain as boundary conditions at 
$\xi \rightarrow -\infty$ the stable state:
\begin{eqnarray} 
b(\xi \rightarrow -\infty ) = c_b, \ \ \ \ \ 
d_{\xi} b(\xi \rightarrow -\infty) =0,  \label{bcb-inf} \\
n(\xi \rightarrow -\infty) = 0, \ \ \ \ \
d_{\xi} n(\xi \rightarrow -\infty) = 0 \label{bcn-inf},
\end{eqnarray}
which invades the unstable state  given at $\xi \rightarrow \infty$:
\begin{eqnarray} 
b(\xi \rightarrow \infty) = 0, \ \ \ \ \
d_{\xi} b(\xi \to \infty) = 0, \label{bcb+inf} \\
n(\xi \rightarrow \infty) =  c_n, \ \ \ \ \
d_{\xi} n(\xi \rightarrow \infty) = 0. \label{bcn+inf}
\end{eqnarray}
As mentioned before, the system simplifies extremely in the region where 
$b(\xi) = 0$.
By choosing the origin $\xi = 0$ in such a way that for positive $\xi$
  $b(\xi) = 0$, the system (\ref{model1db})-(\ref{model1dn}) 
reduces in the positive $\xi$-region to:
\begin{eqnarray} 
b(\xi) &  = & 0, \label{model1db+x} \\
\frac{d^2 n}{d \xi^2} + v_0 \frac{d n}{d \xi} 
&  = & 0.\label{model1dn+x}
\end{eqnarray} 
which is a linear ODE for $n$ which can be solved analytically and is given
by:
\begin{equation} \label{soln+x}
n(\xi) = c_n - c_0 \exp{(-v_0 \xi)},
\end{equation}
where $c_0>0$ is determined by the full problem. 
Hence, the system can be divided into two regimes, the first being 
$\xi > 0$ given by (\ref{model1db+x})-(\ref{model1dn+x}) which can be solved 
analytically, and the second being $\xi < 0$ which contains the full 
nonlinearity. Both 
regimes are connected via their common boundary condition at $\xi = 0$.
Hence, it is sufficient to study (\ref{model1db})-(\ref{model1dn}) for 
$\xi < 0$, for which we still have to determine the behavior at 
$\xi \rightarrow 0$ which we will obtain by studying the local behavior
of the bacteria density $b$ and the nutrient density $n$ as $\xi$ approaches
zero from the left. Since the bacteria density $b$ is a physical quantity, we 
assume it to be continuous. Moreover, (\ref{model1dn}) then implies
that $n$ and 
its derivative at the boundary have to be continous as well.  Hence,
we obtain for $n$ the boundary condition at $\xi = 0$:
\begin{eqnarray}
n(0) & = & c_n - c_0, \label{bcnat0},\\
\left. \frac{d n}{d \xi}\right|_0 & = & v_0 c_0. \label{bcdnat0} 
\end{eqnarray}
In the introduction we have already discussed that the bacteria density $b$
shows a singular behavior for $\xi \rightarrow 0$. This is due to the fact
that the prefactor of the highest derivative in the $b$-equation contains
a factor $b^k$, which vanishes as $b \rightarrow 0$. This allows $b$ to 
become singular near $\xi = 0$. As is well known (see e.g. \cite{Bender})
at such a regular singular point one expects a behavior for $b$ of the type
\cite{note3}:
\begin{equation} \label{ansatz}
b(\xi) = A (-\xi)^\alpha.
\end{equation}
Substituting this Ansatz into (\ref{model1db}) we obtain:
\begin{eqnarray}
& & D \alpha [\alpha(k+1)-1]~ A^{k+1} (-\xi)^{\alpha (k+1) - 2} \\
&  & \hspace{0.7cm} - v_0 A \alpha (-\xi)^{\alpha - 1} - n_0 A (-\xi)^\alpha = 0.\nonumber
\end{eqnarray}
To fulfill this equation the dominant terms in $\xi$, which are
the first and second terms, have to cancel. This determines $A$
and $\alpha$ to be:
\begin{eqnarray}
A & = & \left( \frac{k v_0}{D} \right)^{1/k} , \label{A}\\
\alpha & = & \frac{1}{k}. \label{alpha}
\end{eqnarray}
Hence the bacteria density profile vanishes as
\begin{equation} \label{bcbto0}
b(\xi) \rightarrow \left(- \frac{k v_0}{D}\xi \right)^{1/k} \ \ \ \ 
\mbox{for}\ \ \ \  \xi \rightarrow 0, \label{bto0}
\end{equation}
which implies that its derivative $d b /d\xi$ diverges for $k>1$ as
\begin{equation} \label{bcdbto0}
\frac{d b(\xi)}{d\xi}  \rightarrow -\frac{A}{k} (-\xi)^{1/k - 1} \ \ 
\mbox{for}\ \ \ \  \xi \rightarrow 0.
\end{equation} 
Hence, we are left to study (\ref{model1db})-(\ref{model1dn}) for 
$\xi < 0$ with the boundary conditions (\ref{bcb-inf}), (\ref{bcn-inf}), 
(\ref{bcnat0}), (\ref{bcbto0}) and (\ref{bcdbto0}).

Due to the fact that we chose $f(n,b) = g(n,b)$, a conservation law is
underlying the system (\ref{modelb})-(\ref{modeln}), expressing that all food 
is transformed into bacteria, i.e. that $c_b = c_n$. The conservation law 
allows us to reduce the order of our system of ODE's by one. Hence, by adding
(\ref{model1db}) and (\ref{model1dn}) and integrating over space from 
$-\infty$ to $\xi$ we obtain:
\begin{eqnarray}
\frac{D}{k+1} \frac{d^2 b^{k+1}}{d \xi^2} 
+ v_0 \frac{d b}{d \xi} + n b &  = & 0, \\
\frac{d n}{d \xi} 
+ \frac{D}{k+1} \frac{d b^{k+1}}{d \xi} 
+ v_0 b - v_0 c_b & = & 0. \label{secondeq}
\end{eqnarray}
Note that (\ref{secondeq}) immediately  implies $c_b=c_n$ since the
derivatives all vanish at $\xi \pm \infty$. This just expresses that
food is converted into bacteria in this simplified model.

The one-dimensional front profile governed by (\ref{model1db})-(\ref{model1dn})
 can be represented by a heteroclinic orbit in the $(b,d_{\xi}b,n)$
phase space connecting the two steady states corresponding to the boundary
conditions (\ref{bcb-inf}) to (\ref{bcn+inf}). Due to the possibility of
solving the system of ODE's analytically in the positive $\xi$ region
the front  
profile can be found by applying a standard shooting method to the region 
$\xi < 0$. By shooting from $\xi \rightarrow -\infty$ along the
unstable manifold and requiring it to connect to the trajectory
flowing into the 
singular origin with the boundary conditions (\ref{bcnat0}), (\ref{bcbto0})
and (\ref{bcdbto0}) a relationship between the velocity $v_0$ and the boundary
condition $n(\xi=0)$ is uniquely selected. The existence of a unique front 
solution is also consistent with so-called counting arguments for the
dimensions of the stable and unstable manifolds of the fixed points of
the flow. On the left, 
for $\xi \rightarrow -\infty$, there is only one unstable mode leaving the
homogeneous fixed point, which then fixes $n$ and $d_{\xi} n$ at
$\xi \rightarrow 0$ completely. Matching at $\xi = 0$ to the positive $\xi$
solution for $n$ can only be done on a line in the $n-d_{\xi}n$-plane,
since the $n$-solution is an exponential. Hence, changing $v_0$ so as to match
both fixes $v_0$ completely. 

As we already anticipated at the end of section~I.B, we henceforth
choose $c_n=1$, and hence $c_b=1$: By appropriately rescaling $\xi$,
$v_0$ and the $n$ 
and $b$ fields, any other choice for $c_n$ can be transformed to the
case with $c_n=1$ with renormalized diffusion coefficent $D_R=D c_n^k$. 
The uniquely determined front velocity is hence essentially
a function of $D$ and $k$ only
\begin{equation}
v_0 = v_0(D,k). \label{velo}
\end{equation}

We shall now study the behavior of the front profiles and of $v_0(D,k)$
in more detail by a combination of observations from the numerical
calculations and of simple analytical arguments. Many of these
arguments can easily be formalized by asymptotic analysis or by
reducing the equations in
certain limits to simpler ones, but we shall refrain from doing so
explicitly. 

\begin{figure}[tb]
\begin{center}
\leavevmode
\psfig{figure=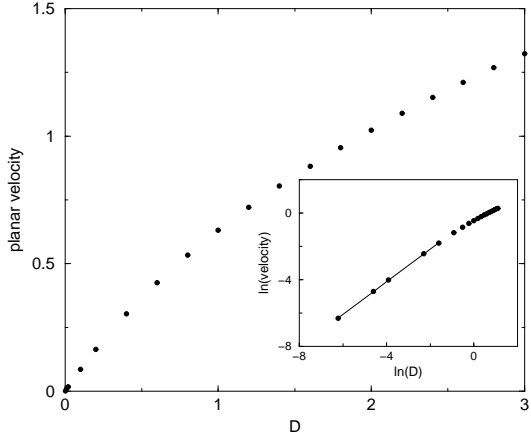,width=7cm}
\end{center}
\narrowtext
\caption{Dependence of the planar velocity $v_0$ on
 $D$ for $k=1$. The inset shows that for 
$D \rightarrow 0$ the velocity approaches zero linearly.}
\label{figlnvelo}
\end{figure}

Fig.~\ref{figvelo} gives an idea of  the functional dependence of the
velocity $v_0$ on  
$D$ and $k$. Fig.~\ref{figlnvelo} displays that for small $D$ the velocity 
is linear in $D$:
\begin{equation}
v_0 \approx a(k) D ~~~~~~~\,\,\,\,\, (D \ll 1),
\end{equation}
where $a(k)$ is a proportionality constant which decreases with increasing $k$.
This proportionality of $v_0$ with $D$ for small $D$ is simply a
consequence of the fact that the propagation of the profile for small
$b$ is governed by the balance of the nonlinear diffusion with the
$v_0db/d\xi$ term. 

\begin{figure}[tb]
\begin{center}
\leavevmode
\psfig{figure=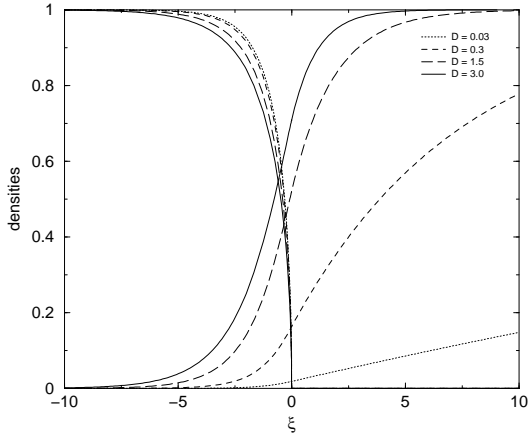,width=7cm}
\end{center}
\narrowtext
\caption{Bacteria and nutrient density profiles for different $D$ and fixed 
$k=2$.}
\label{figDprofiles}
\end{figure}

Fig.~\ref{figDprofiles} shows the dependence of the profile on $D$ for 
$k=2$. 
With decreasing $D$ the interfacial thickness 
$W$ decreases, whereas the diffusion length of the nutrient density $\ell_n$ 
increases since
\begin{equation}
\ell_n = 1/v_0
\end{equation}
as seen from (\ref{soln+x}). Hence, with decreasing $D$ there is a 
separation of  scales between the diffusion length $\ell_n$ and the
interface width.
\begin{figure}[tb]
\begin{center}
\leavevmode
\psfig{figure=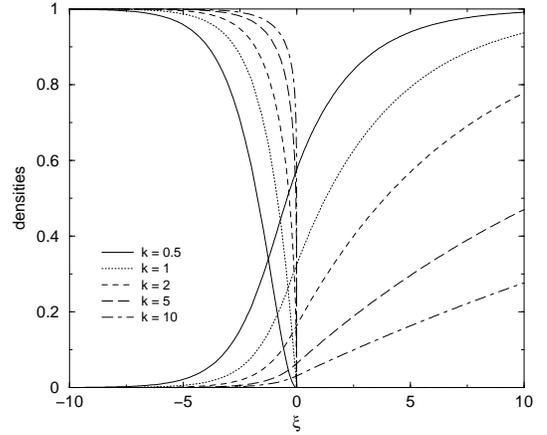,width=7cm}
\end{center}
\narrowtext
\caption{Bacteria and nutrient density profiles for different $k$ and fixed 
$D=0.3$.}
\label{figkprofiles}
\end{figure}
Fig.~\ref{figkprofiles} shows the dependence of the profile on $k$ for 
fixed $D$, here $D=0.3$. 
It demonstrates that with increasing $k$ the interfacial region decreases and
sharpens. 

At first sight, both Figs.~\ref{figDprofiles} and \ref{figkprofiles}
suggest that for small $D$ or large $k$ a moving boundary
approximation might become appropriate. However, the behavior is
rather subtle, and to prepare for a full discussion of this issue in
section~IV, we analyze the scaling of the front profiles in some more detail.

\begin{figure}[tb]
\begin{center}
\leavevmode
\psfig{figure=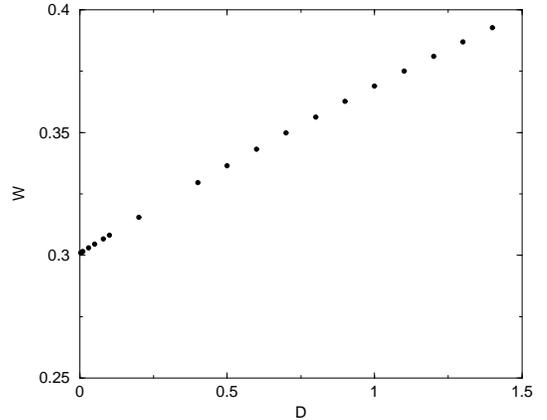,width=7cm}
\end{center}
\narrowtext
\caption{Interfacial thickness as a function of $D$ for fixed $k = 2$.
$W$ is the distance from the origin to the point at which $b=0.5$.}
\label{figDinter}
\end{figure}
\begin{figure}[tb]
\begin{center}
\leavevmode
\psfig{figure=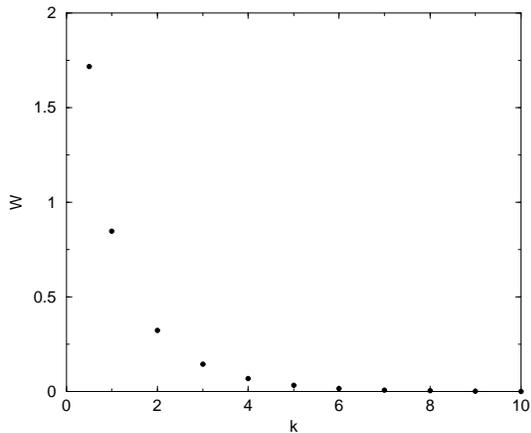,width=7cm}
\end{center}
\narrowtext
\caption{Interfacial position as a function of $k$ and for fixed 
$D=0.3$. $W$ is the distance from the origin to the point at which $b=0.5$.}
\label{figkinter}
\end{figure}
 To quantify the behavior of the interfacial thickness $W$ as a 
function of $k$ and $D$ let us first measure the thickness at which
the bacteria  
density reaches the level $b(W) = b_W = 0.5$. 
Fig~\ref{figDinter} shows how the interfacial thickness approaches a 
finite thickness as $D$ approaches zero, and Fig.~\ref{figkinter} how $W$ 
approaches zero with increasing $k$. Both dependencies can be understood
by inverting (\ref{bto0}):
\begin{equation}
W = \frac{D}{v_0k} b_W^k.
\end{equation}  
Since $v_0$ is proportional to $D$ for small $D$, the interfacial thickness
approaches a constant $W_0$:
\begin{equation}
W \rightarrow W_0 = \frac{1}{a(k) k} b_W^k
\end{equation}
which depends only on $k$ and the chosen interfacial value $b_W$.
With increasing $k$, $W_0$ decreases, and vanishes for $k \rightarrow \infty$;
indeed, for not too large values $b_W$, we have 
\begin{equation}
b^k \approx \exp{( -k |\ln{b_W}|)},
\end{equation}
so that $W_0$ vanishes exponentially. Note finally that Fig.~\ref{figkinter}
indicates that $W$ becomes large as $k \rightarrow 0$; this indicates
that the behavior of the model for $k \ll 1$ is quite different from that in 
the regime $k $ of order 1 or larger, on which we will concentrate.

So far, we analyzed the width between the point where $b$ reaches some
fixed value $b_W <1$ and the point where $b$ vanishes. In the limit $D
\to 0$ this width remains finite, while for $k\to \infty $ the width
measured this way vanishes. However,  for 
addressing the question whether a sharp interface formulation can
capture the essential behavior, it is also important to analyze how
$b$ approaches the asymptotic value 1 for large $k$. When $k$ is
large, we see that 
$n(\xi)$ becomes small in the interfacial zone. In fact, it is easy to
convince oneself that the self-consistent scaling behavior of
Eqs. (\ref{model1db})-(\ref{model1dn}) for $\xi<0$ is $n(\xi) \sim
1/k$, $v_0\sim 1/k$ for large $k$, and this is born out by our numerical
results (not shown). Furthermore, for $v_0$ small and $b\approx 1$,
Eq.~(\ref{model1dn}) for $n$ reduces to $d^2n/d\xi^2 - n=0$, showing
that $n(\xi)$ decays to the left as $e^{-|\xi |}$. In other words, $n$
decays into the bacterial zone on a length scale of order
unity. Through the coupling term in Eq.~(\ref{model1db}), this also
means that $b$ decays to 1  towards the left on a scale of order
unity --- this is actually visible in Fig.~\ref{figkprofiles}. Thus,
even though for large $k$ $b$ rises to values close to 1 on
exponentially small length scales $W$, the scales over which $b$ and $n$
decay to their asymptotic values are actually of order unity.

\section{Linear Stability Analysis of Planar Fronts}
\subsection{Dispersion relation}
To study the linear stability of the planar front, we have to perturb the 
front. Due to the singular behavior of the planar front the
dynamically relevant perturbations
 are not just simply perturbations in the fields $b$ and $n$ but 
also in the shape of the singular  line where $b \rightarrow 0$.
Since we only study the linear stability, we allow the perturbations
to be complex  and we can focus on a single mode with wavenumber $q$
and amplitude $\epsilon$ by writing 
\[h(y,t) = \epsilon \exp{(iqy + \omega t)},\]
We take this function $h$ to be the modulation of the position of the
line where the bacterial front vanishes, as  indicated in Fig.~\ref{figprof3d}.
To be concrete, we now write 
 $b$ and $n$ as
\begin{eqnarray}
b(\xi,y,t)& = & b_0(\xi + h(y,t)) + \nonumber\\
          &   & \epsilon b_1(\xi + h(y,t))\exp{(iqy + \omega t)},\\ 
n(\xi,y,t)& = & n_0(\xi + h(y,t)) + \nonumber \\
          &   & \epsilon n_1(\xi + h(y,t))\exp{(iqy + \omega t)}.
\end{eqnarray}
where $(b_0,n_0)$ is the planar front solution determined in the
previous section.
This ansatz is the {\em crucial ingredient}  that makes our stability analysis 
possible. The standard perturbation approach would amount to writing
the perturbed field $b$ as $b=b_0(\xi)+ \epsilon b_1(\xi)
e^{iqy+\omega t}$; such an Ansatz works only if $b_0(\xi)$ is
smooth enough that its derivative remains finite --- here, because of the singular
behavior of $b_0$, this standard approach fails. We therefore shift
both the position of the singularity line of $b_0$ and of $b_1$, where 
$b_1$ and $n_1$ are the corrections to the bacterial profile and
nutrition field as a result of  this modulation.  In order that
perturbations are arbitrarily small as $\epsilon \to 0$ so that 
we can linearize the equations, we clearly need to have 
\begin{equation}
\frac{n_1}{n_0} ~\mbox{bounded}, ~~~ ~~~
\frac{b_0}{b_1} ~~\mbox{ bounded}. \label{bounded}
\end{equation} Moreover, of course, $b_1$ and $n_1$ should be
continous twice differentiable functions away from the singular
line. 

For the analysis, it will be convenient to introduce the locally co-moving frame
\[ \zeta = x - v_0 t + h(y,t) = \xi + h(y,t) \]
in terms of which the fields can be written as:
\begin{eqnarray}
b(\zeta,y,t)& = & b_0(\zeta) + \epsilon b_1(\zeta)\exp{(iqy + \omega t)},\\ 
n(\zeta,y,t)& = & n_0(\zeta) + \epsilon n_1(\zeta)\exp{(iqy + \omega t)}.
\end{eqnarray}
Upon linearization of the dynamical equations  (\ref{modelb})-(\ref{modeln}) about the uniformly
translating solution $(b_0(\xi),n_0(\xi))$, we then get
\begin{equation}
\mathcal{L} \left( \begin{array}{c} b_1 \\ n_1 \end{array} \right)
= \left( \begin{array}{cc} \omega + \frac{D}{k+1} f^\prime q^2 & 0 \\
                           0 & \omega + q^2 
         \end{array} \right)
  \left( \begin{array}{c} b_1 + \frac{\partial b_0}{\partial \zeta}\\
                          n_1 + \frac{\partial n_0}{\partial \zeta}
         \end{array} \right) \label{linOp}
\end{equation}
where $f=b_0^{k+1}$ and where the prime refers to a differentiation with respect to 
$b_0$. The terms proportional to $\partial b_0/ \partial \zeta$ and
$\partial n_0/ \partial \zeta$ on the right result from the modulation
$h$ 
of the singular line about the line $\xi=0$ in the argument $\zeta$ of
$b_0$ and $n_0$. Finally, the  operator $\mathcal{L}$ is given by
\begin{eqnarray}
{\mathcal{L}}_{11} & = &   \frac{D}{k+1} \frac{\partial^2}{\partial
  \zeta^2} \left( f^\prime \cdot \right)  + v_0\frac{\partial}{\partial
  \zeta} + n_0 
\label{L11}\\
 & = & \frac{D}{k+1} f^\prime 
                        \frac{\partial^2}{\partial \zeta^2} + 
                       \left(2 \frac{D}{k+1} f^{\prime \prime} 
                       \frac{\partial b_0}{\partial \zeta}+ v_0 \right) 
                       \frac{\partial}{\partial \zeta} + \nonumber \\ 
                 &   & \frac{D}{k+1} f^{\prime \prime \prime} 
                       \left(\frac{\partial b_0}{\partial \zeta} \right)^2 + 
                       \frac{D}{k+1} f^{\prime \prime} 
                       \frac{\partial^2 b_0}{\partial \zeta^2} + n_0 ,\nonumber
\\
{\mathcal{L}}_{12} & = & b_0 , \label{L12} \\
{\mathcal{L}}_{21} & = & -n_0 , \label{L21} \\
{\mathcal{L}}_{22} & = & \frac{\partial^2}{\partial \zeta^2} 
                       + v_0 \frac{\partial}{\partial \zeta} - b_0
. \label{L22} 
\end{eqnarray}
Note that the eigenvalue equation (\ref{linOp}) is an ODE problem in
terms of the variable $\zeta$, in the same way as it is in the
standard linear stability calculations \cite{note5}.

Let us pause for a moment to reflect on the difference with the usual
stability approach a bit more.
Since the translational mode $(\partial_{\zeta} b_0,\partial_{\zeta} n_0)$
is the right zero eigenmode of $\mathcal{L}$, 
\begin{equation}
{\mathcal{L}} \left( \begin{array}{c} \frac{\partial b_0}{\partial \zeta} \\ 
\frac{\partial n_0}{\partial \zeta} \end{array} \right) = 0,
\end{equation}
we see that if we  introduce 
\begin{equation}
\bar{b}_1  =  b_1 + \frac{\partial b_0}{\partial \zeta}, ~~~~~~
\bar{n}_1  =  n_1 + \frac{\partial n_0}{\partial \zeta},\label{barbneq}
\end{equation}
in (\ref{linOp}), then these new variables obey simply
\begin{equation}
\mathcal{L} \left( \begin{array}{c} \bar{b}_1 \\ \bar{n}_1 \end{array} \right)
= \left( \begin{array}{cc} \omega + \frac{D}{k+1} f^\prime q^2 & 0 \\
                           0 & \omega + q^2 
         \end{array} \right)
  \left( \begin{array}{c} \bar{b}_1\\
                          \bar{n}_1
         \end{array} \right) \label{linOpbar}
\end{equation}
This is precisely the linear equation one gets if one starts with the usual
linear stability Ansatz $b=b_0(\xi) + \bar{b}_1 (\xi) \exp(iqy+\omega
t)$,  $n=n_0(\xi) + \bar{n}_1 (\xi) \exp(iqy+\omega t)$ in terms of 
$\xi$ rather than $\zeta$  as the variable. While at this
level the two problems appear to be the same, their interpretation is
not. When we write the perturbed problem in terms of the shifted
coordinate $\zeta$ and require $b_1/b_0$ to remain bounded, then
clearly (\ref{barbneq}) shows that the variable $\bar{b}_1$ is {\em
more singular} than $b_0$ --- in particular the singular behavior of
$\bar{b}_1$ is that of $\partial b_0/ \partial \zeta$. In other words,
$\bar{b}_1/b_0$ is  {\em
not}  a small perturbation, instead it diverges! Of course it is simply due to the fact
that one can not represent a shift of the singular line with a small
perturbation in terms of fields which  vanish at $\xi=0$. The Ansatz we make
in terms of the variable $\zeta$, on the other hand, does represent
a proper shift of this line; it can be thought of as a suitable
resummation to capture this.

Let us return to the problem of solving for $b_1(\zeta)$ and $n_1(\zeta)$.
Again we can split up the problem into two separate regions, since for 
$\zeta > 0$ the problem simplifies to:
\begin{eqnarray}
b_1 & = & 0, \\
\frac{\partial^2 n_1}{\partial \zeta^2}  
+ v_0 \frac{\partial n_1}{\partial \zeta} - (\omega + q^2) n_1 
& = & (\omega + q^2) \frac{\partial n_0}{\partial \zeta}.
\end{eqnarray}
which is a linear ODE in $n_1$ which can be solved analytically:
\begin{equation}
n_1 = (c_n-c_0) v_0 \exp{(-v_0 \zeta)} + d_0 \exp{(-\lambda \zeta)},
\end{equation} 
with $\lambda = (v_0-\sqrt{v_0^2+4(\omega + q^2)})/2$ and with $d_0$ some
constant which is undetermined at this stage  ($c_0$ and $c_n$ are
parameters of the solution (\ref{soln+x}) for $n_0$).
This solution is connected to the negative $\zeta$ region via the boundary condition 
at $\zeta = 0$ which determines $d_0$. To obtain the boundary condition at 
$\zeta = 0$, we analyse the local behavior of $b_1$ and $n_1$ as 
$\zeta \rightarrow 0$ from the left. Since $n_1$ and its derivative are 
continuous across $\zeta = 0$, $n_1$ and $\partial_{\zeta} n_1$ obey at 
$\zeta = 0$:
\begin{eqnarray}
n_1 & = & (c_n - c_0) v_0 + d_0, \label{bcn1to0} \\
\partial_{\zeta} n_1 & = & -(c_n - c_0) v_0^2 - d_0 \lambda .\label{bcdn1to0}
\end{eqnarray}  
In view of our requirement (\ref{bounded}) that $b_1/b_0$ remains
bounded, it is natural to assume  that $b_1$ vanishes as
\begin{equation}
b_1 = B (-\zeta)^{\beta}. \label{b1eq}
\end{equation}
Indeed, by inserting it into (\ref{linOp}) we straightforwardly
obtain from the asymptotic behavior (\ref{ansatz}) for $b_0$
\begin{eqnarray}
B & = & -\frac{\omega A}{k v_0} 
    = - \frac{\omega}{k v_0}\left( \frac{k v_0}{D} \right)^{1/k} \label{B} \\
\beta & = & \frac{1}{k}. \label{beta}
\end{eqnarray} 
Hence, for $\zeta \rightarrow 0$:
\begin{equation}
b_1(\zeta) = -\frac{\omega A}{k v_0} (-\zeta)^{1/k} = 
- \frac{\omega}{kv_0} b_0(\zeta), \label{bcb1to0}
\end{equation} 
so that 
\[ \frac{b_1(\zeta)}{b_0(\zeta)} \rightarrow \frac{\omega}{k v} 
\ \ \ \ \mbox{for} \ \ \ \ \xi \rightarrow 0 ,\]
verifying that $b_1/b_0$ remains finite. Hence a solution $b_1$ which
vanishes according to 
(\ref{b1eq}) does obey the requirement that
perturbations are small everywhere.
The boundary conditions at $\zeta \rightarrow -\infty$ are given by:
\begin{eqnarray}
b_1(\zeta) \rightarrow 0, \ \ \ \ \
\partial_{\zeta} b_1(\zeta) \rightarrow 0, \label{bcb1-inf} \\
n_1(\zeta) \rightarrow 0, \ \ \ \ \
\partial_{\zeta} n_1(\zeta) \rightarrow 0, \label{bcn1-inf}
\end{eqnarray}
since all perturbation should vanish at $\zeta \rightarrow -\infty$.

\begin{figure}[tb]
\begin{center}
\leavevmode
\psfig{figure=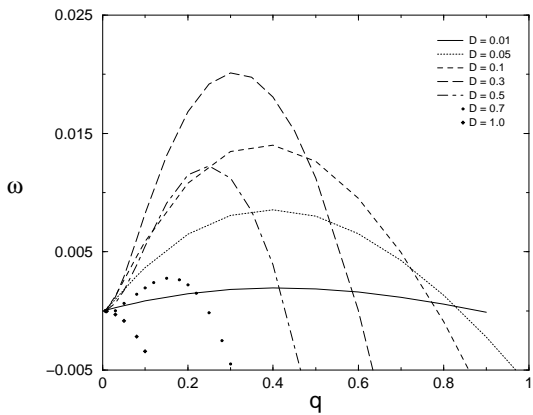,width=7cm}
\end{center}
\narrowtext
\caption{Dispersion relation for $k=2$ and different $D$. For $D<D_c$ the 
planar front is unstable for $q<q_c$ whereas for $D>D_c$ it is linearly stable
for all $q$.}
\label{figdispk2}
\end{figure}

The linear dispersion relation is obtained by solving (\ref{linOp}) for 
different $q$ with the shooting method. By shooting from 
$\zeta \to -\infty$ along the unstable manifold and matching it to 
the trajectory leaving the origin with the boundary conditions 
(\ref{bcn1to0}), (\ref{bcdn1to0}) and (\ref{bcb1to0}) we obtain a unique 
$\omega$ as a function of $D$, $k$ and $q$. At the same time $d_0$ is 
determined. Counting arguments for the multiplicity 
again support the uniqueness of $\omega$.
A numerical dispersion relation was obtained for $k=0.2, 0.3, 0.5, 1, 2, 3$ 
and $5$ and different $D$. For a fixed $k$ the dependence on $D$ of the 
dispersion relation is qualitatively the same for all $k$. 
Fig~\ref{figdispk2} shows the dispersion relation for $k=2$ and different $D$.

There is a long wavelength instability for all $D < D_c(k)$, whereas
all modes are stable for $D>D_c(k)$. As $D$ decreases below $D_c$ the
growth rate of the unstable modes starts to increase as does the range of 
wavenumbers which are unstable. At the same time both $q_m$, the  wave
number which corresponds to the maximum growth rate, as well as $q_c$,
the wave number for which  
$\omega = 0$, shift with decreasing $D$ to larger wave numbers.
By decreasing $D$ even further we observe that the growth rate starts to
decrease again which is due to the fact that the whole dynamics of the front 
is slowing down as we decrease $D$. Note, however, that as $D$ becomes
small, the range of unstable wavenumbers does not vary appreciable:
$q_c$ is roughly constant.
\begin{figure}[tb]
\begin{center}
\leavevmode
\psfig{figure=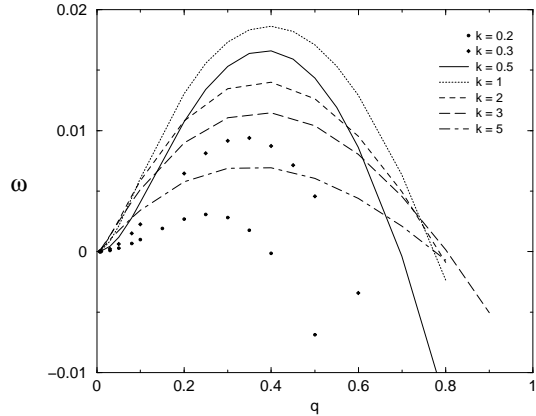,width=7cm}
\end{center}
\narrowtext
\caption{Dispersion relation for $D=0.3$ and different $k$.
}
\label{figdispA}
\end{figure}  
The dependence of the dispersion relation on $k$ is shown in 
Fig.~\ref{figdispA}. We know that for $k=0$ the planar front is stable.
For small $k$ the front starts to be unstable for long wavelength 
perturbations. With increasing $k$ the range of unstable modes is increasing
as is the growth rate. However at $k > 1$ the growth rate starts to decrease
again which is again due to the fact that the whole dynamics of the front
slows down as $k$ is increasing. $q_m$ shows qualitatively
the same behavior as $q_c$. It starts to shift with increasing $k$ to shorter
wave length, stays constant for $k=0.5$ to $k=3$ and then decreases again
to longer wave length.
Fig~\ref{figonset} shows how $D_c$ depends on $k$. With increasing
$k$, the transition value $D_c$
increases, thus implying that with increasing $k$ the region of instability is 
larger. For large $k$ the value of $D_c(k)$ appears to be linear in
$k$. 

 One general noteworthy feature of our results is that the  growth rate of
the most unstable mode as well as the corresponding wavenumber $q_m$
are  generally rather small. As we discussed already in section I, this may the 
reason that Kitsunezaki \cite{kits} appears to observe a planar stable interface
in the region of the phase diagram where planar interfaces are
unstable according to our calculation.

\subsection{Comparison with the Mullins-Sekerka instability}

The dispersion relation of the planar bacterial fronts is, for $D<D_c(k)$ and 
away from
the instability line $D_c(k)$, very similar to the so-called
Mullins-Sekerka dispersion relation
\begin{equation}
\omega_{MS} = v_0 |q| (1 - d_0 \ell_{th}\, q^2) 
\label{msdisp}
\end{equation}
that one derives for perturbations of a planar crystallization
interface \cite{langer}. In this case,  $\ell_{th}=D_{th}/v_0$ is the
thermal diffusion length (the analogue of our nutrient diffusion
length $\ell_n$), and $d_0$ is a microscopic
surface-tension-like length which measures the strength of the
curvature corrections to the interface. We shall see later in section
IV why this analogy is justified, but it already shows us here
something interesting: As $D \to 0$, $v_0 $ vanishes proportional to
$D$. In this limit $\ell_{th}$ diverges just like $\ell_n$
does. Hence, from the observation that the range of unstable modes
remains finite in this limit, and hence that the term analogous to
$d_0 \ell_{th}$ remans finite, we can immediately conclude that the
``effective surface tension'' of our bacterial fronts, the analogue of
$d_0$ in (\ref{msdisp}), should scale as $D$ for small $D$.

\begin{figure}[tb]
\begin{center}
%\leavevmode
\psfig{figure=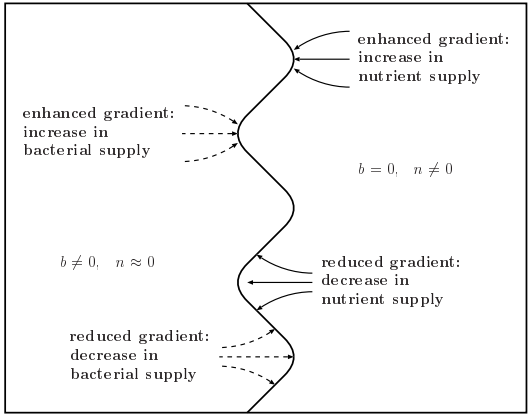,width=7cm}
\end{center}
\narrowtext
\caption{Sketch of a perturbed front propagating from the left to the right.
The arrows drawn with a full line indicate the diffusion flow of
nutrient, on the front side of the interface, those drawn with a
dashed line the diffusion current of bacteria. At a protrusion into
the nutrient region, the nutrient diffusion is enhanced while the
bacterial diffusion current is suppressed. There are hence two
competing effects, whose relative strength depends on $D$. }
\label{figfront}
\end{figure}

That a propagating, planar reaction-diffusion front shows a long wavelength 
instability for small $D$ but is linearly stable for all $q$ for $D > D_c$, 
has been 
observed and explained before (see e.g. \cite{horvath}), and can be
understood in the 
following way. Let us consider a perturbed front moving to the right as 
sketched in Fig.~\ref{figfront}. At a protrusion into the nutrient
side of the interface, the nutrient gradients are compressed and hence 
the nutrient diffusion is enhanced. The ``feeding'' of the interface
from the nutrient side is hence enhanced there, and this tends to make 
such protrusions grow larger in time. On the other hand, as the dashed arrows 
indicate, the bacterial diffusion flow from the back side towards the interface is
reduced at such a protrusion --- this tends to reduce the growth of
such protrusions, and hence to stabilize the interfacial
perturbation. The relative strength of the two effects is determined
by $D$, the effective diffusion constant of the bacteria behind the
interface. When $D>D_c(k)$, the stabilizing effect from the back
side wins, for $D< D_c(k)$  the destabilizing effect on the front
side dominates. 
 Even the effect on $k$ can be
understood in this context. The effective diffusion coefficient is given
by $D = D b^k$, which lowers the effective diffusion coefficient in the 
interfacial region where $b < 1$. Hence, the bigger $k$ the smaller the
effective diffusion constant in the interfacial region, the longer the 
destabilizing effect of the nutrient can prevail.  When $k$ decreases
towards zero, the stabilizing bacterial diffusion extends more and
more towards the front region \cite{notek0}.

As we pointed out above , in the limit $D\ll D_c(k)$ the instability is
very much like the classical Mullins-Sekerka instability of a
crystal-melt interface. As $D$ increases towards $D_c$ this connection 
breaks down because the stabilizing diffusion from the back-side
becomes important {\em within}   the interfacial zone: There is then no clear 
separation anymore between an interface and the regions before and
behind the front [see also section IV.C for further discussion of the
behavior for $D$ near $D_c(k)$].

Of course, the competition between the stabilizing effect of the
diffusion gradient on the back side and the destabilizing effect of the
gradient on the front side of the interface shows up in crystal growth 
during transient regimes and can be understood along the lines of the
 Mullins-Sekerka stability analysis \cite{langer}. A most amusing and dramatic
illustration of this was observed recently in experiments on the
melting of polarized $^3He$ \cite{melting}; there the instability sets
in only after 
a very long transient because the diffusion coeffient on the back side
is very much bigger than on the front side; as a result, as long as
there is a transient gradient on the back side, the melting interface remains stable.

\subsection{Onset of instability}
As we found above that the instability that occurs when $D $ decreases
below $D_c(k)$  is a long-wavelength $q=0$ instability, the critical
line  $D=D_c(k)$ is the line where 
 $d\omega / d(q)^2|_{q=0} =0 $: to the right of this line in
 Fig.~\ref{figonset} this derivative is negative and to the left of
 it it is positive, so that $\omega >0 $ for small $q$.
 Since the translational mode 
$q=0$ is the  eigenmode of $\mathcal L$ with eigenvalue $\omega = 0$, we can  
investigate the behavior of the $\omega$-$q^2$ curve in the vicinity of the 
origin by the following expansion \cite{toth}. Because the $q=0$ 
mode is a translation mode with zero eigenvalue, $\omega$ is small and 
of order $q^2$ when $q$ is small. Moreover, $b_1=0$ and $n_1=0$ for
$q=0$, and so for small $q$, $b_1$ and $n_1$ are both of order $q^2$ too.
In (\ref{linOp}) this implies that 
for $q$ small, the terms on the right hand side involving $b_1$ and
$n_1$ are of order $q^4$. To order  $q^2$, we therefore get 
\begin{equation}
\mathcal{L} \left( \begin{array}{c} b_1 \\ n_1 \end{array} 
\right)
= \left( \begin{array}{cc} \omega + \frac{D}{k+1} f^\prime q^2 & 0 \\
                                                 0 & \omega + q^2
         \end{array} \right)
  \left( \begin{array}{c} \frac{\partial b_0}{\partial \zeta}\\
                          \frac{\partial n_0}{\partial \zeta} 
  \end{array} \right) , \label{onsetlinOp}
\end{equation}
which is exact to order $q^2$. 
 Since $\mathcal{L}$
has a zero eigenvalue, we can apply the solvability condition by
requiring that the inner product with the left zero mode of
$\mathcal{L}$ vanishes:
\begin{equation}
\int_{-\infty}^\infty d\zeta \left( \begin{array}{c} \Psi_1 \\ 
                                                     \Psi_2 
                                    \end{array} \right)^T
      \left( \begin{array}{cc} \omega 
             + \frac{D}{k+1} f^\prime q^2 & 0 \\ 0 & \omega + q^2
             \end{array} \right)
      \left( \begin{array}{c} \frac{\partial b_0}{\partial \zeta}\\
                          \frac{\partial n_0}{\partial \zeta} 
             \end{array} \right) = 0, \label{onsetSolv}
\end{equation}
Here $\Psi_1$ and $\Psi_2$ are the components of the left zero mode,
i.e., of the right zero eigenvector 
of the adjoint matrix operator $\mathcal{L^*}$:
\begin{equation}
{\mathcal{L^*}} = \left( \begin{array}{cc}
                  \frac{D}{k+1} f^\prime \frac{\partial^2}{\partial \zeta^2} 
                  - v_0 \frac{\partial}{\partial \zeta} + n_0     & -n_0 \\
                  b_0 & \frac{\partial^2}{\partial \zeta^2} 
                  - v_0 \frac{\partial}{\partial \zeta} - b_0 \end{array}
                  \right). \label{onsetAdj}
\end{equation}
Upon rewriting (\ref{onsetSolv}) as
\begin{eqnarray}
&  \omega & \int_{-\infty}^\infty d\zeta \left( 
        \Psi_1 \frac{\partial b_0}{\partial \zeta} + 
        \Psi_2 \frac{\partial n_0}{\partial \zeta} \right) \nonumber \\
& = & - q^2 \int_{-\infty}^\infty d\zeta \left( 
      \frac{D}{k+1} \Psi_1 f^\prime \frac{\partial b_0}{\partial \zeta} +
      \Psi_2 \frac{\partial n_0}{\partial \zeta} \right),
\end{eqnarray}
and taking the limit $q^2 \to 0$, this
leads to the required exact relation for the onset of the lateral instability:
\begin{equation}
\left. \frac{d \omega}{d (q^2)} \right|_{q=0} =
 - \frac{\int_{-\infty}^\infty d\zeta \left( 
         \frac{D}{k+1} \Psi_1 f^\prime \frac{\partial b_0}{\partial \zeta} +
        \Psi_2 \frac{\partial n_0}{\partial \zeta} \right) }
{\int_{-\infty}^\infty d\zeta \left( 
        \Psi_1 \frac{\partial b_0}{\partial \zeta} + 
        \Psi_2 \frac{\partial n_0}{\partial \zeta} \right)}
\ . \label{onsetCon}
\end{equation}
Planar fronts change  stability when the integral in the numerator of (\ref{onsetCon})
changes sign. 

 Since $\mathcal{L}$
is non-hermitean, there is no obvious relationship between the zero right eigenmode
of $\mathcal{L}$ and its adjoint $\mathcal{L^*}$. To find the zero right 
eigenvector of the adjoint operator $\mathcal{L^*}$ we have to impose
appropriate boundary conditions on the left eigenmodes too. Generally,
the boundary conditions of the functions in the left adjoint space
are obtained from 
the definition of the adjoint operator, in that for all functions $\Phi$ we
consider we need to have
\begin{equation}
\int_{-\infty}^\infty d\zeta \Psi ({\mathcal{L}} \Phi) 
= \int_{-\infty}^\infty d\zeta ({\mathcal{L^*}} \Psi) \Phi.\label{innerproddef}
\end{equation}
In general, when the partial integrations are done so as to obtain
$\mathcal{L}^*$ from $\mathcal{L}$, we obtain boundary terms; the
requirement that these vanish give the boundary conditions on the
adjoint functions $\Psi$. In the
present case, since the functions on which our operators are working
are defined on the infinite interval $(-\infty,\infty)$ for $n_1$  and
its related left 
component $\Psi_2$, and on the semi-infinite interval $(-\infty,0]$ for 
$b_1$ and $\Psi_1$, we find that the appropriate boundary condition
for the adjoint functions $\Psi$ is that $\Psi_2$ should stay bounded
 as $\pm \infty$; likewise $\Psi_1$ should stay bounded both as 
 as $\zeta \to -\infty$ and as $\zeta\to 0$.

We are now in a position to  analyze the behavior of the adjoint
eigenmodes; we will report the analysis in some detail, as the various
elements form  important ingredients of the derivation of a moving
boundary approximation in the next section. 
$\mathcal{L^*}$ simplifies again considerably in the positive $\zeta$
region due to the fact that $b_0$ vanishes  identically there, so it
is again of 
advantage to split the region of integration into two, $\zeta < 0$ with 
$\mathcal{L^*}$ given by (\ref{onsetAdj}), and $\zeta > 0$ for which
\begin{equation}
{\mathcal{L^*}} = \left( \begin{array}{cc}
                  - v_0 \frac{\partial}{\partial \zeta} + n_0     & -n_0 \\
                  0 & \frac{\partial^2}{\partial \zeta^2} 
                  - v_0 \frac{\partial}{\partial \zeta} \end{array}
                  \right). \label{onsetAdjx+}
\end{equation}
For $\zeta > 0$ $\Psi_2$ has to solve the homogeneous ODE:
\begin{equation}
\frac{\partial^2 \Psi_2}{\partial \zeta^2} 
- v_0 \frac{\partial \Psi_2}{\partial \zeta} = 0.
\end{equation}
This equation has two independent solutions, a constant and an
exponential that diverges for increasing $\zeta$. Hence the boundary
condition that $\Psi_2$ remains bounded immediate gives the solution
\begin{equation}
\Psi_2 = \psi_0 = \mbox{constant, }~~\zeta>0. \label{bcPsi2to0}
\end{equation}
Moreover, the differential equations for $\Psi_2$ implied by
the zero-eigenvalue equation
\begin{equation}
{\mathcal{L}}^* \left( \begin{array}{c}  \Psi_1 \\ \Psi_2 \end{array}
  \right) =0  \label{lefteigenvalue}
\end{equation}
 shows that $\Psi_2$ has to be continuous and
have a continuous derivative at $\zeta=0$. Hence, when we construct
the eigenmodes on the left half space $\zeta <0$, the $\Psi_2$ component has to
obey the boundary condition
\begin{equation}
\Psi_2 (\zeta=0) = \psi_0, ~~~~~\partial \Psi_2 / \partial \zeta
|_{\zeta=0}=0.\label{bcat0}
\end{equation}

Since $b_0$ vanishes identically for $\zeta > 0$, we need to know $\Psi_1$
only in the region $\zeta < 0$. As we stated above, because the
functions $b_1$ that we consider all vanish as $\zeta\uparrow 0$, the
definition of the adjoint operator does not imply a boundary condition
on $\Psi_1(\zeta=0)$ as long as it does not diverge. A
straightforward analytical investigation of the equation near
$\zeta=0$ shows that in general $\Psi_1$ will, with a finite slope,
approach a finite value as $\zeta \uparrow 0$, and that in general it has a
higher order singular term $\sim (-\zeta)^{(1+D/k)}$.

We now turn to the behavior as  $\zeta \rightarrow -\infty$. In this
limit, $n_0\to 0$ and $b_0\to 1$, so $\mathcal{L}^*$ reduced from
(\ref{onsetAdj}) to
\begin{equation}
{\mathcal{L^*}} = \left( \begin{array}{cc}
                  {D}  \frac{\partial^2}{\partial \zeta^2} 
                  - v_0 \frac{\partial}{\partial \zeta}     & 0 \\
                  1 & \frac{\partial^2}{\partial \zeta^2} 
                  - v_0 \frac{\partial}{\partial \zeta} - 1 \end{array}
                  \right). \label{onsetAdj2}
\end{equation} 
It is easy to verify that as $\zeta \to -\infty$, there are three
possible types of non-divergent zero mode solutions,
\begin{equation}
\Psi^{(1)} \sim \left( \begin{array}{c} 1  \\ 1
\end{array}\right) ,~~~~
\Psi^{(2)} \sim e^{v_0\zeta/D},~~~~\Psi^{(3)}\sim e^{\lambda_+\zeta}. \label{psis}
\end{equation}
Here $\lambda_\pm = (v_0\pm \sqrt{v^2_0+4})/2$, so that the mode $\sim
e^{\lambda_+\zeta}$ indeed converges towards the left; the other mode
allowed by the linear equations, $e^{\lambda_-\zeta}$, on the other
hand, diverges towards the left, and hence is forbidden by the
boundary conditions. 

The mode $\Psi^{(1)} $ is very special --- it is immediately verified
from (\ref{onsetAdj}) that 
\begin{equation}
{\mathcal{L}^*} \left( \begin{array}{c} \psi_0 \\ \psi_0
\end{array}\right) =0 ~~\mbox{for all }\zeta ~~~~(\psi_0 ~\mbox{const.}),
\end{equation}
not just for $\zeta\to -\infty$. In other words, the constant mode
$\Psi^{(1)}$ is an exact adjoint zero mode for all $\zeta \le 0$.

If we integrate  $\Psi^{(2)}$ or $\Psi^{(3)}$ forward towards
increasing $\zeta$, each trajectory in the phase space of the ODE is
uniquely determined (apart from an overal amplitude, as the equations
are linear). Hence, if we follow either $\Psi^{(2)} $ or $\Psi^{(3)}$
towards $\zeta=0$, the derivatives $\partial_\zeta
\Psi^{(2)}|_{\zeta=0}$ and  $\partial_\zeta
\Psi^{(3)}|_{\zeta=0}$ will in general be {\em nonzero}. As we have
however seen above, in order that the full eigenmodes on the whole
real axis remain bounded also for $\zeta\to \infty$, the
$\Psi_2$-component needs to have a zero derivative at $\zeta=0$ ---
see Eq. (\ref{bcat0}). Each separate eigenmode does not obey this
requirement, but by the linearity of the equation it will always be
possible to construct one unique linear combination $\bar{\Psi}^{(2)}$
of $\Psi^{(2)}$ 
and $\Psi^{(3)}$ which does have zero derivative. This solution
constitutes the second adjoint zero eigenmode of the problem. Like the
trivial eigenmode $\Psi^{(1)}$, it can be extended smoothly to the
required $\bar{\Psi}^{(2)}_2=const.$ behavior for $\zeta>0$. 

\begin{figure}[tb]
\begin{center}
\leavevmode
\psfig{figure=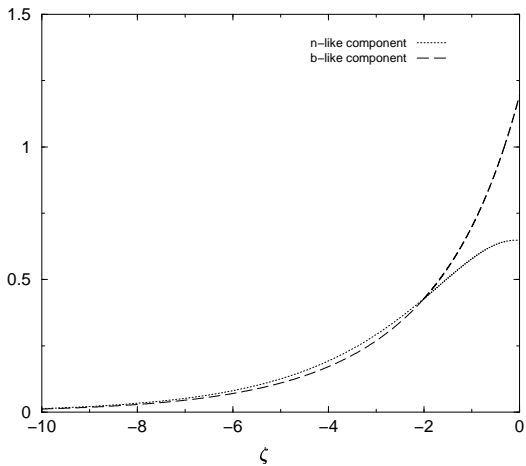,width=7cm}
\end{center}
\narrowtext
\caption{Zero right eigenvector $\bar{\Psi}^{(2)}$ of the adjoint
operator ${\mathcal{L}}^*$ for 
$k=2$ and $D=0.3$. The $b$-like component $\Psi_1$ approaches a finite
value at $\zeta=0$ with a finite slope; it has generally a higher
order singular 
term $\sim (-\zeta)^{(1+D/k) }$.}
\label{figleft}
\end{figure}
Fig.~\ref{figleft} shows the adjoint zero eigenmode
$\bar{\Psi}^{(2)}$  of $\mathcal{L^*}$ for $k=2$
and $D=0.3$, obtained numerically from the ODE's with a shooting
method. The qualitative behavior of the components  is in agreement
with the above discussion, and is independent of the values of the
parameters. Note that the $b$-like component $\bar{\Psi}^{(2)}_1$
approaches a finite value at $\zeta=0$ with a finite slope; this
behavior is also found for arbitrary parameters, while as is easily
verified there generally is a parameter-dependent subdominant singular
term
proportional to  $|\zeta|^{(1+D/k)}$. 
 
To obtain the onset of lateral instability the adjoint zero eigenmode
has to be convoluted with the translational mode 
$(\partial_\zeta b_0,\partial_\zeta n_0)$ according to
(\ref{onsetCon}). Which of the two zero modes should we use? 
The trivial adjoint zero mode $\Psi^{(1)}$ expresses change of
velocity under reparametrization, and is 
related  to conservation law in  our system 
[see the discussion after Eq. (\ref{secondeq})]; this
will become more clear in the next section. It does not play a role
for the onset of instability; we will therefore ignore it here
\cite{note4} and use $\bar{\Psi}^{(2)}$ to evaluate (\ref{onsetCon}). 
A change of sign of the numerator, which marks  the onset of
instability, is obtained 
for $k = 0.5, 1, 2$ and $3$ and shown in Fig.~\ref{figonset} as
diamonds. In the figure, these values are
 also compared to the $D_c$ determined by the numerical dispersion 
relation shown in Fig.~\ref{figonset} as filled dots. 
The agreement is very good, as it should; we have also checked that
away from this line, a fit of the small-$q$ behavior of the dispersion 
relation leads to values of the second derivative of $\omega$ at $q=0$ 
which are consistent with the solvability formula. These results thus
confirm the 
consistency of our full stability calculation and the solvability
expression for the critical line in the phase diagram and the small
$q$ behavior of the growth rate $\omega$.

\section{Sharp interface formulation}
A moving boundary approximation or sharp interface formulation is
appropriate when the 
width of the front or interface is much smaller than the typical length scale
of the pattern and when the dynamics of the pattern occurs through the motion
of these interfaces. The moving-boundary approximation amounts then to treating
these fronts as mathematically sharp interfaces or boundaries by taking their 
width to zero and integrating out their internal degrees of freedom. 
There are three important assumptions underlying such an
approximation, namely {\em (a)} that there is a separation of length scales, 
{\em (b)} that there is a separation of time scales between the motion of the front 
as a whole and its internal dynamics, and {\em (c)} that the internal dynamics of
the front is determined by the nonlinear front region itself, so that the
solvability-type integrals are dominated by the contributions from the finite
region, and hence do not diverge.  The latter condition is violated
 in 
practice only for special types of fronts propagating into a linearly
unstable state, so-called ``pulled'' fronts \cite{ute2,ute1}; our fronts
are not of this type (they are of the  ``pushed'' type, in this
terminology),  so we focus our analysis on the length and time scale
requirements {\em (a)} and {\em (b)}.

As we saw in section III, the planar front width is finite, even for
$D\to 0$ at fixed $k$. Moreover, even though for $k\to \infty$ the
$b$-field rises over an exponentially small distance behind the
singularity line, both the $b$ and $n$ field even then only
approach their asymptotic values over a distance of order unity. In
this sense, even in this limit the front width $W$ remains finite. Of
course, we can always {\em choose} to investigate fronts whose
curvature $\kappa$ is small in the sense that $\kappa W \ll 1$. For
these, a moving boundary approximation should be accurate; we do find
indeed below that the sharp interface formulation we derive for the
present problem is consistent with the result of the dispersion
relation of section III for $q$ small enough that $qW\ll 1$.  However, 
whether such a moving boundary approximation applies {\em at all
dynamically relevant length scales}, is another matter! We already
know from the analysis of the dispersion relation in section III that
in the left part of the $D$-$k$-phase diagram modes up to $q=q_c$ are
unstable, and that $q_c$ is generally  finite, except close to the
critical line $D=D_c(k)$. Hence, $q_cW$ remains finite as well, and so 
there is no obvious limit where a moving boundary approximation
becomes asymptotically correct on all dynamically relevant length
scales. Nevertheless, we find that in practice the sharp interface
formulation that we develop is rather accurate in a significant portion of the 
phase diagram. Since the present problem has some unusual and new
aspects, we focus again on the essential structure and intuitive
arguments, rather than mathematical rigor.

\subsection{Sharp Interface Formulation of the Problem}

The simplest case to consider to guide our intuition is the limit
$D \ll 1$. As we discussed in connection with
Fig.~\ref{figDprofiles}, in this regime the bacterial density field
approaches its 
asymptotic value on the finite scale $W$, while the $n$-diffusion
field in front of the bacterial 
front decays on a length scales $l_n=1/v_0$, which diverges as $D\to 0$
since $v_0\sim D$. A sharp interface
formulation is then based on the idea that we view the bacterial front
on the ``outer'' scale $\ell_{out}$ on which the patterns and
diffusion fields vary in the presence of the moving boudary,
which is treated as a sharp line of zero thickness. The dynamics of the fields
on the ``inner'' scale \cite{fife,vandijk} of the front ($W$), and the way in which this
dynamics responds to changes in the fields on both sides of this inner 
zone, is translated into boundary conditions for the outer fields in
the interfacial formulation. Formally, the moving boundary consists of
taking the limit $\delta \equiv W/\ell_{out} \to 0$. 

Normally, a sharp
interface formulation or moving boundary approximation is based on
applying the theory of matched asymptotic expansions
\cite{Bender,vandijk}. Here, the situation is somewhat unusual: on the
right side of the inner region (the interfacial transition zone where
essentially the nutrient is consumed by the bacteria) we
already have a sharply defined boundary, the singular line where $b$
vanishes. At this side, we therefore do not have a matching problem,
instead we  already have two boundary conditions for the
$n$ field, namely the requirements that the value of $n$ and the
gradient of $n$ are continous at the singular line where $b$ vanishes --- this follows directly from the
dynamical equation (\ref{modeln}), since the ``reaction term'' $-nb$
is continuous. On the left side of the interfacial
zone, on the other hand, the $b$ field varies continuously and we do
have a true matching problem. 

In our present case, the ``outer field'' on the front side of the
interface is simply the $n$ field in the whole region to the right of
the singular line where $b$ vanishes; hence there the nutrient field $n$
obeys a simple linear diffusion equation:
\begin{equation}
\label{nouter}
\mbox{front side ``outer'' eqs.:} \left\{ \begin{array}{l}  b=0, \\ \\
\frac{\partial n}{\partial
t} = \nabla^2 n. \end{array} \right.
\end{equation}
It is useful to introduce a suitable curvilinear coordinate system in
which $\xi=0$ coincides with the singular line where $b$ vanishes. In
the sharp interface limit, the line $\xi=0$ then also coincides with
the position of the moving interface. We furthermore identify the
region ahead of the front as the + side of the interface where
$\xi>0$, and use a superscript + to indicate values of the outer field
extrapolated from the outer region towards the line $\xi=0$: $n^+=
\lim_{\xi\downarrow 0} n(\xi)$,
 $\nabla n^+ = \lim_{\xi\downarrow 0} \nabla n(\xi)$. As we
already mentioned, $n$ and its gradient should be continuous at this
line, and we can therefore  write the boundary conditions as
\begin{equation}
 \lim_{\xi\uparrow 0} n(\xi) = n^+,~~~~~~\lim_{\xi\uparrow 0} \nabla
n(\xi) = \nabla n^+. \label{matchbcn}
\end{equation}

We now turn to the behavior on the back side of the front, the $-$ side.
We have seen that in the bacterial front,
the nutrition field decays exponentially fast to zero towards the left
on a length scale 
of order unity; hence, in the $-$ outer region behind this interfacial zone,
we have $n\approx 0$ for the nutrition field. The bacterial field  is
close to 1 there. Therefore, we 
take the dynamics of the $b$-field into account there, by writing
$b=1+\Delta b$ and linearizing 
in the outer field $\Delta b$,
\begin{equation}
\label{bouter}
\mbox{back side ``outer'' eqs.:} \left\{ \begin{array}{l} \frac{\partial \Delta
b}{\partial t} = D \nabla^2 \; \Delta b, \\ \\ n=0. \end{array} \right.
\end{equation}
Note that the outer equations (\ref{nouter}) and (\ref{bouter}) have
been written in the laboratory frame, not in a co-moving frame, since
in the case of nontrivial patterns, there is no single relevant
co-moving frame. 

What are the boundary conditions on the $-$ side of the front? According
to the matching prescription, {\em the inner field expanded in the outer
variables on the back side should equal the outer field expanded in
the inner variables} \cite{vandijk}. Extrapolating the inner field in the outer
variables towards the $-$ side means that we investigate the $b$-profile to
the left towards $-\infty$. As Figs.~\ref{figDprofiles} and
\ref{figkprofiles} illustrate, on the inner scale $W$ the $b$ field
rapidly approaches a constant value. Although these figures are made
for planar fronts, the analysis below  shows
that this continues to hold for weakly curved fronts and 
that the appropriate matching conditions are
\begin{equation}
\mbox{matching condition:} \left\{ \begin{array}{l} \lim_{\xi\to -\infty} b (\xi) = 1+ \Delta
b^-, \\ \\ \lim_{\xi\to -\infty} \nabla b (\xi)=0. \end{array} \right. \label{matchbcb}
\end{equation}
Here $\Delta b^-$ is the value of the outer field $\Delta b$
extrapolated from the outer $-$ region towards the interface. Note that
the second condition that the gradient vanishes, is also consistent
with the matching prescription: if we assume that the outer field
$\Delta b$ varies on the outer scale ${\bf X} =\delta {\bf r}$ with
$\delta = W /\ell_{out} $ then  the outer
gradient of $\Delta b$ rewritten in terms of the inner variable
vanishes in the limit $\delta \to 0$.

Now that we know how to connect the inner fields to the outer ones ---
on the left side of the inner region through the matching conditions (\ref{matchbcb}), on
the right side through  boundary conditions
(\ref{matchbcn}) --- we are ready
to derive the effective boundary conditions in a sharp-interface
formulation. One easily convinces oneself that in order to get a
well-posed moving boundary problem with the above outer dynamical
equations and matching conditions, one needs effectively {\em three}
boundary conditions relating   $\Delta b^-$, $n^+$ and $\nabla n^+$
and the interface velocity and curvature.  To derive them,
  we imagine that the front is weakly curved with
curvature $\kappa$ such that $\kappa W  \propto \delta \ll 1$.
Since we identified the line $\xi=0$ with the line where $b$ vanishes,
$\xi$ is a local co-moving  coordinate  with speed $v$
  in the direction perpendicular to the front. In this weakly curved
 system, we then have  on the inner scale
\begin{eqnarray}
\frac{\partial b }{\partial t} & = &  
\frac{{D}}{k+1} \frac{ \partial^2 b^{k+1}}{\partial \xi^2} + \left( v + D \kappa \, b^k \right) \frac{\partial b}{\partial \xi}  
 + {n} b,  \label{curvedeqb}\\
 \frac{\partial {n}}{\partial t } & = &  
\frac{ \partial^2 {n}}{\partial  \xi^2} +\left( {v} + \kappa \right) \frac{\partial {n}}{\partial \xi} 
- {n} b . \label{curvedeqn} 
\end{eqnarray}
Following standard practice, we now ignore the time derivatives on the 
left (taken in the co-moving frame). This amounts to an adiabaticity
assumption, the assumption
that the change in the pattern and hence in the front speed and
profile, take place on time scales much longer than the relaxation
time of the front (assumption {\em (b)} discussed in section IV.A
above). Technically, it means 
that  the solution stays always close to a uniformly
translating solution in the curved coordinate system, and our goal now 
is to calculate the changes in the velocity perturbatively. Indeed,
let us write
 $v=v_0 +v_1$, where $v_0$ is the velocity of the planer
solution and $v_1$ the change in velocity due to the curvature and the 
fact that the outer field $n$ is changed slightly from the planar
solution; similarly we write $b=b_0 + b_1^\prime$ and
$n=n_0+n_1^\prime$ so that $b_1^\prime$ and $n_1^\prime$ are the
deviations  
in the fields from the planer solutions (we use the prime to emphasize 
the difference from the perturbations used in the linear stability
analysis). Upon linearization about the planar front solution, we
then get the equation
\begin{equation}
\mathcal{L} \left( \begin{array}{c} b_1^\prime \\ {n}_1^\prime \end{array} 
\right)
= \left(\begin{array}{cc} -{v}_1 - D \kappa\, b_0^k & 0 
\\
                                                 0 & -v_1-\kappa
  \end{array} \right)
  \left( \begin{array}{c} \frac{\partial b_0}{\partial \xi}\\
                          \frac{\partial {n}_0}{\partial \xi} 
  \end{array} \right), \label{curvedop}
\end{equation}
where $\mathcal{L}$ is the same operator introduced before in
(\ref{linOp}), written now  in terms of the variable $\xi$. 
This equation again calls for applying the solvability condition. We
have already seen in section III.C that the operator $\mathcal{L}$ has 
a number of  adjoint zero eigenmodes.  There is a
subtle difference between the present analysis and the one of section III.C,
however, which 
is crucial for obtaining the proper boundary conditions. In the
stability analysis of section
III.C we worked with functions defined on the whole interval
$(-\infty,\infty)$; this  implied that the mode constructed for
negative $\zeta$ needed to have zero derivative of the $\Psi_2$
component at $\zeta=0$, and this reduced the number of proper adjoint
zero eigenmodes  to two. Here, however, we are doing perturbation
analysis only in the inner region, which in the inner variable  is the
semi-infinite interval
$(-\infty,0]$. We therefore do not need to require now that the
adjoint zero mode has $\partial \Psi_2/ \partial \xi|_{\xi=0}=0$, and
 hence we now have {\em three} rather than two admissable adjoint zero
modes! These lead to the three necessary equations that become the
boundary conditions sought for in the sharp interface formulation.
Moreover, because we now work on a semi-infinite interval, we  get
boundary terms at $\xi=0$  from the partial
integrations necessary to have the operator work on the adjoint
functions \cite{note9}:
\begin{eqnarray}
\lefteqn{\int_{-\infty}^0  d\xi\; \left( \Psi_1,\Psi_2\right) \cdot {\mathcal{L}}
\left( \begin{array}{c} b_1^\prime \\ n_1^\prime \end{array} \right)}\nonumber
\\
&  &~~~~= 
\int_{-\infty}^0 d\xi \left[ {\mathcal{L^*}} \left( \begin{array}{c}
      \Psi_1 \\ \Psi_2 \end{array} \right) \right]^T \cdot
\left( \begin{array}{c} b_1^\prime \\ n_1^\prime \end{array} \right)\nonumber\\
& & ~~~~-\Psi_1(-\infty) \, v_0 \Delta b^-   -
\left. \frac{\partial \Psi_2}{\partial \xi}\right|_{\xi=0} (n^+
-n^+_{0} ) \nonumber \\ & & ~~~~ + \Psi_2 (0) \left[ 
\nabla_{\xi}n^+ -\nabla_{\xi} n^+_{0}  + v_0 (n^+ -n^+_0 ) \right]
. \label{adjointeq}
\end{eqnarray}
Here we have used the boundary and matching conditions
(\ref{matchbcb}) and (\ref{matchbcn}) for the deviations
$b_1^\prime$ and $n_1^\prime$ from the planar values $b_0$ and $n_0$.
Note that there are no boundary terms in the
field $b_1^\prime$ at $\xi=0$, since these are all proportional
to $b_0^k $, and $b^k_0(\xi\to 0)$ vanishes according to
(\ref{bcbto0}).

The three boundary conditions now straightforwardly follow by taking
the left inner product of the three zero modes with  equation
(\ref{curvedop}) together with 
(\ref{adjointeq}). The behavior of the three adjoint zero  modes of
${\mathcal{L}}^*$ on the left for $\xi \le 0$ 
has been discussed already in section III.C. The first one is simply
$\Psi^{(1)}_1=\Psi^{(1)}_2=constant$. In this case we immediately obtain
\begin{eqnarray}
\lefteqn{  - v_0 \Delta b^-  + ( \nabla_\xi n^+ -  \nabla_\xi n^+_0 ) + v_0( n^+ - n^+_0)}\nonumber \\ & &~~= - \int_{-\infty}^0 d\xi \left( [ v_1+D \kappa
b_0^k] \frac{\partial b_0}{\partial \xi} + [v_1+\kappa ]
\frac{\partial n_0}{\partial \xi} \right)\nonumber \\
 & & ~~=  v_1 [1 -n_0^+]  + \kappa [ D/(k+1) - n_0^+]. \label{firstbc}
\end{eqnarray}
We note that this equation is essentially a type of conservation
equation in a weakly curved frame --- indeed, it can also be obtained
by an analysis similar to the derivation of the conservation equation
(\ref{secondeq}) by adding the two equations (\ref{curvedeqb}), and
(\ref{curvedeqn}) and integrating, ignoring the temporal derivates for
a quasi-stationary front solution in the co-moving frame. This is the
reason for our earlier remark in section III.C that the constant left zero mode
$\Psi^{(1)}$ is related to conservation. 

The second and third boundary conditions are  obtained from the two
other adjoint  zero modes
$\Psi^{(2)}$ and $\Psi^{(3)}$ of
$\mathcal{L}^*$, discussed in section III.C; the general form is
similar to the one shown for a $\bar{\Psi}^{(2)}$ for a 
particular choice of parameters in Fig.~\ref{figleft}, except that the
derivative of the $n$-like component does not vanish at $\xi=0$. These
zero eigenmodes of $\mathcal{L}^*$ can only be evaluated numerically,
but the form of the boundary condition  is simply the same for both:
with $i=2,3$ we get
\begin{eqnarray}
\lefteqn{   \Psi^{(i)}_2(0)\left( (\nabla_\xi n^+ -\nabla_\xi
n^+_0 \right) } \nonumber \\ & & ~~~~~~~~~~+  \left( v_0 \Psi^{(i)}_2 -\left. \frac{\partial \Psi_2^{(i)}}{\partial \xi}\right|_0\right) (n^+-n^+_0 )  \nonumber \\ & &~~~~= -
\int_{-\infty}^0 d\xi\; \Psi^{(i)}_1 \left( [ v_1+D \kappa
b_0^k] \frac{\partial b_0}{\partial \xi} \right) \nonumber\\
 & & ~~~~~~~~~~-
\int_{- \infty}^0 d\xi\; \Psi^{(i)}_2
 [v_1+\kappa ]
\frac{\partial n_0}{\partial \xi} \nonumber \\
& &~~~~ = - v_1 \int_{-\infty}^0 d\xi\; \left( \Psi^{(i)}_1
\frac{\partial b_0}{\partial \xi}  + \Psi^{(i)}_2 \frac{\partial
n_0}{\partial \xi} \right) \nonumber \\ & &~~~~~~~~~ - \kappa
\int_{-\infty}^0 d\xi\; \left( \Psi^{(i)}_1 D b_0^k \frac{\partial
b_0}{\partial \xi}  + \Psi^{(i)}_2 \frac{\partial n_0}{\partial \xi}
\right). \label{secondbc} 
\end{eqnarray}
In the sharp interface interpretation, equations (\ref{firstbc}) and
(\ref{secondbc}) are interpreted as the boundary conditions that relate the
change in the local field $n_1^\prime$ and its gradient  at the
interface (relative to
those for a planar moving front) to the local change in velocity of
the interface and the local curvature. For a general pattern, the
derivative of $n_1^\prime$ with respect to $\xi$ on the left of these
equations has to be interpreted as derivative in the direction normal
to the interface\cite{note7}. As we discussed above, these equations are
precisely the three boundary conditions necessary to get, together with
the outer equation (\ref{nouter}), a well-posed moving boundary
problem. 

\subsection{Interpretation of the sharp interface formulation}
It is useful to pause for a moment to reflect on the structure of the
boundary conditions. First of all, note that they all involve terms
proportional to the gradient of the nutrient diffusion field, and not
to the bacterial density field. The presence of these gradients of the
nutrient field on the front side of the interface could have been
expected from the numerical observations that the bacterial growth
fronts are unstable for small enough $D$. It is well know
\cite{langer}
that such
interfacial instability arise for diffusion limited growth problems
where
 the growth velocity of the
interface is proportional to the gradient of the driving field that
``feeds'' the interface. The fact that the gradient of the bacterial
field $b$ does not appear, makes these bacterial growth fronts most
like the so-called ``one-sided crystal growth models'', describing situations
where the diffusion on the back side is absent (e.g., in ``directional
solidification'', the diffusion of impurities in the solid on the back
side of the interface is usually negligible in comparison with the
diffusion in the liquid on the front side \cite{langer}). 

We can make
this observation more precise as follows. Note that the boundary
condition (\ref{firstbc}), the one that expresses conservation,  is
the only one which involves $\Delta b^-$. Hence we can solve the full
dynamical problem by working {\em only} with the outer $n$-field and the two
boundary conditions (\ref{secondbc}) --- in other words, the outer
$n$-field together with these boundary conditions constitute a closed
problem that is sufficient to describe the dynamics of the moving
interface. Once this is determined, one can use the conservation
condition (\ref{firstbc}) to determine $\Delta b^-$ and from there
analyze the behavior of the $b$-field on the back with the outer
equation (\ref{bouter}): {\em in the sharp
interface limit the $b$-field
becomes completely slaved to the interface motion!}

We already anticipated in section III.B that the curvature correction
term (the effective surface-tension-like term) should be of order $D$
for small $D$. This is fully confirmed by our analysis: all the
curvature terms in the boundary conditions (\ref{firstbc}) and
(\ref{secondbc}) either explicitly involve a term $D$, or a term
proportional to $n_0$ which, as we saw in section II, is proportional
to $D$ too for small $D$. 

\subsection{Applicability to the various regimes}
The simplest way to test the accuracy of a moving boundary formulation
is by comparing  the dispersion relation from the 
moving boundary problem with the dispersion relation obtained from the
full model as discussed in section III.A. The two outer  equations are
linear diffusion equations of the standard form, while the boundary
conditions are (by construction) also linear. Consequenty, the
stability of the planar 
solution of the sharp interface problem follows the standard stability
problem as discussed in \cite{langer} in which the growth or decay of
small single-mode perturbations about the planar interface solution is
determined.   We will therefore not
report it explicitly here.
\begin{figure}[tb]
\begin{center}
\leavevmode
\psfig{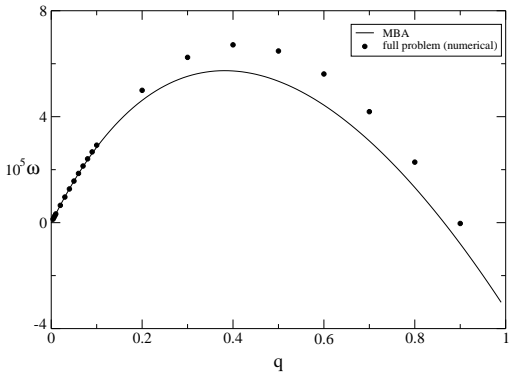}
\end{center}
\narrowtext
\caption{Comparison between dispersion relation obtained through a sharp
interface approach (full line) and through direct numerical linear
stability analysis (dots), for the case $k=2$, $D=0.001$. Note that
for $q\ll1$ the results from the moving boundary approximation  agree
very well with those of the full linear stability calculation, as it
should. The reason for the difference between the two curves for $q$
values of order unity is discussed in the text.} 
\label{figsharpA}
\end{figure}
In Fig.~\ref{figsharpA} we illustrate such a comparison for a typical
case in the small $D$ regime, where the moving boundary approximation
is expected to work best since the diffusion length in the nutrient +
region. The  figure, which is for the case   $k=2$ and $D=0.001$
confirms that for small $q$ the dispersion relation of the moving
boundary approximation (full curve) essentially lies on top of the one derived from
the full problem (symbols). This is as it should, since for small $q$ clearly
$qW\ll 1$, so that the condition for the moving boundary approximation to
be accurate is fullfilled. The overal shape of the dispersion relation
of the moving boundary approximation is actually quite close to the
exact one, but for larger $q$ there clearly are some quantitative
differences, even for this small value of $D$. This discrepancy is in
our view due to what we discussed before, the fact that the range of
unstable wavenumbers for this case is finite ($q_c \approx 0.8$),
while the interface width $W$ is finite too, so that $q_cW $ does not
approach zero as $D\to 0$.

Even though the moving boundary
approximation therefore does not become formally correct in this limit
(in the sense that the correction terms can not be made arbitrarily
small by taking $D$ sufficiently small), it clearly does quite well in
practice for these parameter values. Probably this is due to the fact
that $W$ is still relatively small compared to the wavelength
$\lambda_c= 2\pi / q_c$ 
corresponding with the marginal wavenumber $q_c$: If we take $W\approx
2$  we get $W/ \lambda_c \approx 1/4$, so even though we can not make
this ratio arbitrarily small by sending $D\to 0$, it appears to be
small enough in practice to make the sharp interface formulation work
well.  What may also play a role is 
that for problems with nonlinear diffusion like this one, the response
of the interfacial zone to perturbations is mostly determined by the rather
thin zone where $b$ is small; we have not attempted to substantiate
this intuitive idea, however.

A similar observation holds for the timescales. As
Fig.~\ref{figdispk2} illustrates, the maximum
growth rate $\omega_m$ of the most unstable mode is proportional $D$
and hence $v_0$ for
small $D$; the proportionality $\omega_m \simeq v_0 q_m$ is also
consistent with the Mullins-Sekerka 
dispersion relation (\ref{msdisp}) discussed in section III.A. The
internal relaxation time $\tau_{front}$ of the front itself is expected to be of
order $W/v$, hence $\omega_m T \simeq q_m W$ remains finite in the limit $D \to
0$: there is no full separation of timescales either.

Also for  $k\gg 1$ and $D$ of order unity, the present approximation
works generally very
well, since on the one hand the diffusion length in the nutrient zone
ahead of the front is large (as $v_0\sim 1/k$ for $k$ large), while on
the other hand the interfacial zone tends to becomes relatively small,
even though it does not appear to go to zero --- see
Fig.~\ref{figkprofiles} and the discussion at the end of section
II. We have also investigated the possibility whether in the limit 
$k\gg1$ another approximation might be possible, one in which there
are three zones, an outer region in front of the interface where $b=0$
as we had above, a very thin zone (of exponentially small width, see
section II) where $b$ quickly rises to a value close to 1 while $n$
hardly changes, and a region behind this zone where $\Delta b= b-1$ is
already
small and where $n$ decays to zero. We have not been able to make this
approximation work to our satisfaction, basically because we have not
been able to match the thin zone properly to the region behind it. 

For values $D$ and $k$ of order unity, but not too close to the
stability boundary $D_c(k)$, the discrepancy between dispersion
relation of the moving
boundary approximation that we have derived above and the exact
dispersion relation is bigger than in Fig.~\ref{figsharpA} for small
$D$. This is to 
be expected, since in this limit diffusion length in the outer
$n$-region is of the same order as the interface width, and, moreover,
the diffusion in the bacterial region is more important. Nevertheless, the
order of magnitude of the growth rate and the range of unstable
wavenumbers is right.

Finally, we note that since a
long-wavelength instability occurs upon decreasing $D$ below $D_c(k)$,
we expect that just to the left of this line, the dynamics can be
described by the so-called Kuramoto-Sivashinsky equation
\cite{kuramoto,kapral}.  In
fact, since the line $D_c(k)$ is very straight for $k $ larger than 1, it
may be well be the problem also simplifies in the limit $D\to \infty$,
$k \to \infty$, $D/k $ fixed.
We have not attempted to study this limit or to give an explicit
derivation of the Kuramoto-Sivashinksky equation near the boundary.

\section{Summary and outlook}
We have shown that a nonlinear diffusion coefficient and a simple
bilinear autocatalysis is sufficient to generate a long wave length 
instability as long as the diffusion constant obeys $D < D_c$, where $D_c$
depends on the nonlinearity.  We hope that these results will be of help in sorting out
to what extent the present class of models does describe the real bacterial growth problems. To
do so, one would of course have to be able to map the bacterial growth properties onto
the effective diffusion coefficient in this model. If this can be done, the most clear test with the
aid of the present results would be to see whether the interfacial instability becomes suppressed
once the effective bacterial density becomes too large.  

It would also be of interest to extend numerical simulations like those of  Golding {\em et al.} \cite{benjacob3} and those of Kitsunezaki \cite{kits},
whose parameter values are indicated by crosses in Fig.~\ref{figonset}: As we discussed in
the introduction, these  numerical results appear to contradict our analytical results, but this
may be due to finite size effects and the limited time of the simulation.

In addition to providing a starting
point for further studies of these models for bacterial growth, we
have been able to develop a new type of  linear
stability calculation which applies to the general class of fronts
with singular behavior of the fields as a result of nonlinear
diffusion. Our methods therefore opens up the possibility to study
other systems as well, like the vortex fronts
\cite{gilchrist,surdeanu}
we mentioned in the
introduction. 
In addition we were able to reformulate the reaction diffusion
problem with nonlinear diffusion
away from the stability boundary $D_c(k)$  in the form of a free
boundary type problem; our analysis shows that in the present model
bacterial growth fronts
are closest to those described by the so-called one-sided model in the
theory of crystal growth \cite{langer}. 

 Of course, our results are far from the   final answer 
on these type bacterial growth models: the knowledge that
the planar front  
is unstable is only the first (though crucial) step towards
understanding   the actual evolving 
patterns, which are determined by nonlinear effects. In addition,
within the context of understanding the bacterial growth problem, the
question remains to what extent models with nonlinear diffusion
suffice to capture the important growth dynamics.

Clearly, we have studied only  the simplest variant of such models, by
leaving out the death term, which appears to be important for the
morphology \cite{kits,benjacob}, as well as  lots of effects
which are important for a more realistic model for bacterial colony
growth, like the sporulation of bacteria 
 already mentioned in the introduction:  Computer simulations
have shown that in order that branches can form, the sporulation term
$- \mu b$ has to be present in (\ref{modelb}).

We have not tried to study how the critical line $D_c(k)$  approaches
the origin as $k\downarrow 0$; this is an interesting technical
question, but one which probably is of limited relevance for
understanding the bacterial growth problem. In fact,  the
behavior near the origin in the $D$-$k$-phase diagram is very singular 
and hence
sensitive to changes in the model: the nonlinear diffusion as well as
finite cutoff effects as well as changeing the bilinear reaction term
$bn$ to $b^{\gamma}n$ change the mathematical behavior dramatically.

Finally, we want to draw attention to an open mathematical question
--- at least for us. In a solvability type analysis, the boundary
conditions one normally imposes on the adjoint fields follow from the
requirement that $\langle \Psi ({\mathcal{L}}\Phi)\rangle = \langle
({\mathcal{L^*}}\Psi)\Phi\rangle$ for all  dynamically relevant
functions $\Phi$ [see Eq. (\ref{innerproddef})]. However, in the
derivation of our moving boundary approximation, we have operated
differently: instead of imposing boundary conditions on the adjoint
functions, we have written out the terms from the partial differential
equations explicitly, and used the left zero modes on the half space
$(-\infty,0]$ that we already knew to obtain the sought for boundary 
conditions for the physical fields $n_1$! Clearly, the equations 
obtained this way follow necessarily from the original differential
equations in the weakly curved frame, but this line of reasoning is
mathematically different in spirit from the usual Fredholm alternative
(solvability theory). We do not know  --- nor could we find --- the 
mathematical theory behind this approach which appears to be new  and  
very powerful for problems with a singular line.   

\section{Acknowledgement}
The work by JM in Leiden was made possible through a postdoctoral position within the EU network "Patterns, Noise and Chaos". 

% results and discussion
 
% \vspace{-.8cm}

\end{multicols}

\begin{references}
\bibitem{note1} In technical terms: the fronts in coupled reaction
diffusion equations with a linear bacterial diffusion term and a
bacterial growth term which is linear in the bacterial density, have so-called
``pulled'' planer fronts which are not unstable. Introducing a small
density cutoff as proposed in \cite{kessler} makes such fronts weakly
``pushed'' and unstable; introduction of a nonlinear diffusion
coefficient of the type proposed by Ben-Jabob and co-workers makes the
fronts strongly pushed and also unstable.
\bibitem{note2a} For $\gamma=1$ and linear diffusion, the fronts are
  ``pulled'', meaning that their asymptotic velocity equals the
  spreading speed $v^*$ of the linearized equations. These fronts are
  known to be very special, in that perturbation theory does not apply 
  to them (see \cite{ute2}). For any $\gamma >1$ but linear diffusion, 
  the linear spreading speed $v^*=0$, and hence fronts in this case
  have to be pushed. Standard perturbation theory and arguments then
  apply to them. See \cite{ute2} for further discussion of this point.
\bibitem{note3} We note that the system (\ref{modelb}), (\ref{modeln})
also admits 
    solutions with $b$ decreasing exponentially as $b \propto \exp{(-\xi/v_0)}$
    for $\xi \rightarrow \infty$. The existence of such additional solution
    is quite common for fronts propagating into an unstable state 
    \cite{Bender}. Typically, such solutions are dynamically irrelevant for 
    initial conditions which are sufficiant localized (e.g. have compact 
    support). The simulations of \cite{kits} indicate that this is
    also the case here, and we will therefore not investigate such solutions
    any further.
\bibitem{note5} Although the function appearing in the stability
calculations are effectively only functions of one variable, $\zeta$,
we prefer to use partial derivatives in this section to stress that we
do consider perturbations that both depend on time and on the
$y$-variable.
\bibitem{notek0} The issue of why fronts are always stable for $k=0$ is 
  quite subtle --- compare Ref. \onlinecite{note1}. For $k=0$ the
  fronts are pulled, and then the semi-infinite region ahead of the
  front determines the dynamics of a front, rather than the nonlinear
  dynamics within the front region itself.  We refer to
  Ref. \onlinecite{ute2} for further discussion of this point.
\bibitem{note4} The solvability condition for this mode yields $q=0$,
  while $\omega$ is undetermined;
hence this mode corresponds to changes in the planer solution. As we
have seen in section II, changes in the constants $c_b$ and $c_n$ for
the asymptotic behavior of $b$ and $n$ yield to planar front solutions 
with a different velocity. In view of this and the fact (discussed in
section IV) that this mode is associated with the conservation law, we
assume that this mode does not  
play a role for perturbations that leave the asymptotic values of
$c_b$ and $c_n$ unchanged.
\bibitem{note6} This is because for $\xi>0$ $\Psi_2$ of this mode is a
constant, and because the $n$-component of this zero eigenmode is
continuous and has a continuous derivative at $\xi=0$.
\bibitem{note9} Since the analysis used to obtain the
boundary conditions  from the partial integration terms is somewhat
nonstandard we stress that the equations derived by taking the inner
products with the three adjoint zero modes are clearly necessary
conditions. As pointed out also section V, how the analysis compares
to the standard Fredholm 
alternative (solvability condition) is not quite clear to us.
\bibitem{note7} It is instructive to realize that when we consider a
slight shift in 
the position of a planar front, for the outer field $\partial 
n_1^\prime /\partial \xi |_0 + v_0n_1^\prime =0$, as follows 
immediately from the outer field solution (\ref{soln+x}) for the planar 
case. Thus, the terms on the left hand side of the above boundary 
conditions both vanish, and hence both $v_1=0$ and $\kappa=0$, as it 
should for a shift of a planar front solution.
\end{references}
\end{document}